\documentclass[12pt]{article}

\usepackage{cite}
\usepackage{amsmath,amsfonts,amssymb}
\usepackage{graphicx}
\usepackage{psfrag}
\usepackage{enumerate}
\usepackage{epsfig}
\usepackage{color}
\usepackage{enumerate}

\makeatletter \@addtoreset{equation}{section}

\makeatletter\renewcommand\section{\@startsection {section}{1}{\z@}%
                                   {-3.5ex \@plus -1ex \@minus -.2ex}
                                   {2.3ex \@plus.2ex}%
                                   {\normalfont\large\bfseries}}
\renewcommand\subsection{\@startsection{subsection}{2}{\z@}%
                                     {-3.25ex\@plus -1ex \@minus -.2ex}%
                                     {1.5ex \@plus .2ex}%
                                     {\normalfont\bfseries}}

\newcommand{\be}{\begin{equation}}
\newcommand{\ee}{\end{equation}}
\newcommand{\beq}{\begin{eqnarray}}
\newcommand{\eeq}{\end{eqnarray}}

\def\[{\left [}
\def\]{\right ]}
\def\({\left (}
\def\){\right )}

\def\r2{\sqrt{2}}

\newcommand{\bbibitem}[1]{\bibitem{#1}\marginpar{#1}}

\def\Label#1{\label{#1}%
  \smash{\hbox to0pt{\raise1ex\hbox{\tiny[#1]}\hss}}}
\def\noLabels{\let\Label=\label}
\def\nobbibitem{\let\bbibitem=\bibitem}

\begin{document}
\noLabels 
\nobbibitem 

\begin{titlepage}


\vfil\

\begin{center}

{\Large \bf Quantitative approaches to} \\
{\Large \bf information recovery from black holes}

\vspace{3mm}

Vijay Balasubramanian$^{a,}$\footnote{\tt email:vijay@physics.upenn.edu},
Bart{\l}omiej Czech$^{b,}$\footnote{\tt email:czech@phas.ubc.ca}

\vspace{5mm}

\bigskip\centerline{$^a$\it David Rittenhouse Laboratory, University of Pennsylvania,} \smallskip\centerline{\it 209 South 33$^{\rm rd}$ Street, Philadelphia, PA 19104, USA }
\bigskip\centerline{$^b$\it Department of Physics and Astronomy,  University of British Columbia,}
\smallskip\centerline{\it 6224 Agricultural Road, Vancouver, BC V6T 1Z1, Canada}

\vfil

\end{center}
\setcounter{footnote}{0}
\begin{abstract}
\noindent
The evaporation of black holes into apparently thermal radiation poses a serious conundrum for theoretical physics:  at face value, it appears that in the presence of a black hole quantum evolution is non-unitary and destroys information.  This information loss paradox has its seed in the presence of a horizon causally separating the interior and asymptotic regions in a black hole spacetime.   A quantitative resolution of the paradox could take several forms: (a) a precise argument that the underlying quantum theory is unitary, and that information loss must be an artifact of approximations in the derivation of black hole evaporation, (b) an explicit construction showing how information can be recovered by the asymptotic observer, (c) a demonstration that the causal disconnection of the black hole interior from infinity is an artifact of the semiclassical approximation.   This review summarizes progress on all these fronts.
\end{abstract}
\vspace{0.5in}

\end{titlepage}
\renewcommand{\baselinestretch}{1.05}  


\tableofcontents

\section{Introduction}

In 1974 Hawking analyzed quantum mechanical fields propagating in the background of matter collapsing to form a black hole \cite{Hawking:1974rv, Hawking:1974sw}. He found that the incipient horizon affects field  modes by mixing positive and negative frequencies in such a manner that at late times thermal radiation emerges from the black hole.   At least in the semiclassical approximation in which these calculations are done, the radiation at each time is in an exactly thermal density matrix, and leads to slow evaporation of the black hole.  Eventually, the black hole disappears, its energy dissipated to infinity by the radiation.   This leads to a puzzle.  We could have formed the black hole initially from a pure state of matter. But the final state created by radiation in a thermal density matrix appears to be oblivious to all details of the initial state, except for its mass, angular momentum, and global charges.  Thus it is not possible for the asymptotic observer to determine the state of the matter that originally formed the black hole. If this conclusion is correct, then quantum mechanics in the presence of a black hole is non-unitary \cite{Hawking:1976ra}, threatening its consistency as a physical theory.  

There is a related puzzle arising from the large entropy associated to a black hole.  Consider, for example, an eternal black hole that exists for all of time in equilibrium with a bath of radiation \cite{Hawking:1982dh} or a very large black hole that is evaporating very slowly. Such a black hole is thought to have an entropy proportional to its horizon area: $S = A/4G_N \hbar$ \cite{Bekenstein:1973ur}. In any consistent quantum theory of gravity we expect  this entropy to be explained by the existence of $e^S$ microstates that are commensurate with the macroscopic parameters of the black hole (mass, angular momentum, and global charges).    Microcanonically, the black hole can be in any one of these microstates.  However, the classical observer outside the black hole has no way of determining this microstate, and thus should interact with the black hole as if it were in a density matrix over the underlying states.  Indeed, correlation functions computed by an asymptotic observer in the presence of a classical black hole horizon decay at late times as they would in a thermal background \cite{KeskiVakkuri:1998nw}. Is there any method, even in principle, of  ``detecting a black hole microstate'', or is that information always lost to the asymptotic observer?

The most mysterious property of a black hole, and the one that gives it its name, is the fact that the metric can be smoothly continued past the horizon giving a geometric region that extends all the way to a spacetime singularity.  The interior region is causally disconnected from infinity -- observers following inertial trajectories can fall in unmolested, but cannot make their way back out because all causal trajectories in the interior point towards the singularity.    Thus the doings of the infalling observer, and other events that occur inside the horizon, seem to be inaccessible to the outside.   Tracing over the interior degrees of freedom again leads the outside observer to lose information about the quantum state of the universe.  There is a subtle point that the initial conditions for all infalling modes can be set outside the horizon, and hence one could imagine that all the information is somehow also represented outside the black hole.   To make matters even more confusing, in the time coordinates of an observer who remains outside the black hole, nothing ever falls behind the horizon -- all infalling matter gets redshifted and piles up at the horizon.  While this raises the question of whether the interior ``exists'' in a meaningful sense for external observers, na\"\i vely tracing over the interior will cause the external observer to represent the black hole as a density matrix.  

The preceding paragraphs relate the information loss problem to three questions: (a) what are the properties of the vacuum state in the presence of a horizon? (b) can an asymptotic observer identify a black hole microstate? (c) can the fate of infalling observers be detected at infinity? These questions highlight an important conceptual point, which suggests an avenue for reconciling black holes with unitarity. If the area of the black hole horizon is related to its entropy, how should we think of a black hole that is in a pure microstate, a single member of the underlying statistical ensemble? After all, a pure state has no microscropic entropy except in a coarse grained description. Thus, the relation between horizon area and entropy mandates that a universe in a pure black hole microstate must not have a wavefunction localized on a geometry with a finite area horizon. The latter might emerge after some coarse-graining, but the microscopic picture must be fundamentally different. Perhaps it does not even make sense to talk about the interior of a black hole except, at best, for some kinds of coarse grained observers. In this picture, the causally disconnected black hole interior would be an artifact of the semiclassical approximation and there would simply be no room for information loss.

This might sound like an unacceptably drastic departure from the conventional perspective. On the other hand, {\it all} approaches to avoiding information loss have to invoke some exotic escape from na\"\i ve semiclassical reasoning. For example, it has been argued \cite{Lowe:1995pu} 
that interactions between the enormously blueshifted infalling quanta and the enormously redshifted quanta climbing out of a black hole involve very high center of mass energies, which suggests that low energy effective field theory cannot be used near the horizon. This claim has engendered two approaches to the black hole information problem. The idea of complementarity \cite{Susskind:1993if} states that since the temporal redshift at the horizon prevents asymptotic observers from witnessing anything fall into a black hole, asymptotic and infalling observers give equivalent descriptions of the same physics. Meanwhile, Maldacena proposed that quantum gravity enjoys a dual ``holographic'' description in terms of degrees of freedom of a lower dimensional field theory \cite{Maldacena:1997re}, which is manifestly unitary. The approach championed in this paper makes extensive use of holographic techniques. These examples illustrate that solving the information problem requires one to be skeptical about semiclassical intuitions. Perhaps the least radical detour from the traditional picture would be to restore unitarity with subtle correlations in Hawking radiation, but as we discuss in the text, it is difficult to produce the requisite correlations sufficiently quickly.

This review article emphasizes that the restoration of unitarity involves resolving corrections to the classical results which are of magnitude $\mathcal{O}(e^{-S})= \mathcal{O}(e^{-A/4G_N \hbar})$.    In terms of scaling with $\hbar$ and $G_N$ the corrections are evidently not perturbative and become immeasurably small as $\hbar \to 0$, explaining the loss of information in the semiclassical limit.   We will further argue that the scale $\mathcal{O}(e^{-S})$ arises from {\it statistics}, not dynamics, and that quantum gravity mainly enters the problem in determining the non-perturbatively tiny gap in the spectrum that is necessary to account for the enormous entropy of black holes.   In making these arguments, it will frequently be convenient to talk about extremal black holes, and especially those where the horizon is of zero size and coincides with the singularity.   These objects are not the conventional black holes that result from gravitational collapse, but they provide a useful laboratory. They are stable, because their temperature vanishes, and they often have a substantial degeneracy, though not generally enough to result in a finite horizon.  External probes of these black holes shed light on the conceptual foundations of information recovery from black holes.  It will also be convenient to consider black holes in Anti-de Sitter (AdS) space, because in this case we can expect to describe the physics of spacetime using the dual conformal field theory \cite{Maldacena:1997re, Witten:1998qj}. In these settings, there is simply no room for information loss, because the dual field theory is manifestly unitary. This allows us to concentrate on a more precise question: what is the mechanism for the recovery of information from black holes?

\section{General remarks}
\label{preliminaries}

\subsection{Virtues and features of a laboratory in AdS space}
\label{bhads}

Much of the progress in thinking about black holes and information has been in the context of asymptotically AdS spacetimes.  Below we will describe such black holes and explain why they are such a useful laboratory.    Any resolution of the information paradox that works in AdS backgrounds should in principle carry over to all black holes.  This is because  Hawking's argument for information loss arose from the structure of local quantum field theory in the vicinity of the horizon and does not depend in an essential way on the asymptotics of the spacetime. 
On the other hand, the global properties of AdS make it a particularly advantageous laboratory for studying the information paradox. This section reviews the many virtues of AdS: it is convenient for calculations and it affords a sharper definition of the information paradox, in which its structure (as well as its resolution) become more apparent. These advantages are traced to the AdS/CFT duality \cite{Maldacena:1997re, Witten:1998qj}, which posits that gravity in a spacetime asymptotic to AdS space enjoys an equivalent formulation in terms of a conformal field theory on the boundary of AdS. For a review of AdS/CFT, the reader is referred to \cite{Aharony:1999ti}.


Black holes in AdS fall into two classes, small and large \cite{Hawking:1982dh}. Small black holes have horizon radii, which are small compared to the AdS curvature radius and as a result of that their physics is qualitatively similar to black holes in flat space. Large AdS black holes, on the other hand, are qualitatively different.
First, their specific heat is positive, so they are stable minima in the canonical ensemble. Second, because AdS acts effectively as a box, large black holes can be \emph{eternal}, that is they can attain thermal equilibrium with their own, reflected Hawking radiation.

These features are convenient for studying the information problem. Consider dropping a quantum into an eternal black hole. There are two sharply defined scenarios for the subsequent evolution of the system:
\begin{enumerate}
\item Deviations from the thermal description of the black hole decay to zero and information is lost.
\item Deviations from the thermal description initially decay as the black hole churns the fallen information, but eventually they hover around some small finite value and information is preserved.
\end{enumerate}
The facts that large AdS black holes are stationary and well-described in the canonical ensemble are essential for defining and contrasting the two scenarios.


In fact, the AdS/CFT correspondence \cite{Maldacena:1997re} automatically eliminates the non-unitary scenario, because the dual conformal field theory is manifestly unitary.  (The only subtlety is that it may be necessary to include separate components of the conformal field theory on causally disconnected segments of the AdS boundary.) This explains away the information paradox in its basic form, but it does not make clear the precise way in which Hawking's original argument fails. Before declaring victory over the information paradox, one must first provide a satisfactory account of where and how Hawking's derivation breaks down. This is the essence of the AdS/CFT-motivated version of the information problem.

Before launching an attack on the information paradox in AdS, it is useful to contrast the ingredients of holographic duality with those of Hawking's argument.
The AdS/CFT correspondence, which from the viewpoint of string theory is a manifestation of open-closed string duality, arises when we consider the near-horizon limit of a large stack of parallel D-branes. In the low energy regime the asymptotic and near-horizon regions decouple, and one identifies the low energy dynamics of the D-branes with string theory on the near-horizon geometry, which always contains an AdS factor. The duality is of the weak-strong type: weakly coupled gravity in AdS maps into strongly coupled world-volume field theory. Furthermore, the duality interchanges the IR with the UV: local disturbances in the field theory correspond to gravity deformations on a very large scale and near the boundary of spacetime \cite{Susskind:1998dq}. Thus, though the dual CFT provides in principle a complete definition of quantum gravity of AdS, its manifest unitarity comes at the expense of having a simple description of local physics in spacetime. In particular, it is difficult to use the CFT description to track how information falls into and leaks out of a black hole.

On the other hand, Hawking's derivation uses methods of local effective field theory, with local quantum fields propagating on a fixed classical spacetime containing a horizon. Na\"\i vely, this argument should be valid so long as the Schwarzschild radius of the black hole is large in Planck units, because in that regime the calculation is safe from effects of quantum fluctuations.  At least some of the approaches presented in this review challenge the assumption that the concept of a fixed, classical black hole spacetime remains valid in the vicinity of the horizon. 


The information paradox depends on another important problem in quantum gravity -- that of explaining the entropy of a black hole. Information can only be preserved if the internal state (microstate) of a black hole changes uniquely after any specific external state falls in.   If one could detect differences between internal states, there would be no information paradox.  Conversely, to resolve the information paradox is tantamount to demonstrating that there exist microstates, which are in principle distinguishable, though not necessarily to semiclassical observers.   In addition, there should be some microscopic sense in which each microstate is horizonless because a pure state has zero entropy.  From this perspective, the causal disconnection of the black hole interior, which results in the thermality of black hole radiation, should be demonstrably a semiclassical arifact.

The AdS/CFT correspondence provides a framework in which these words may be translated into actual computations. Different states in asymptotically AdS spacetime map into different states in the dual field theory: empty AdS corresponds to its ground state, AdS with some particles corresponds to a low energy excited state, and a black hole corresponds to a thermal state. If the arguments of the previous paragraph are correct, then the black hole geometry with a horizon could be seen to arise as an effective, thermodynamic description of many (microscopically horizonless) microstates. In gravity it is not clear how to define such microstates and how to enact a thermodynamic averaging over them, but in field theory this is natural. The dual CFT contains a notion of an inner product on its Hilbert space, so one can work with an orthonormal basis of states. This basis is discrete, because the curvature of AdS breaks translational invariance and produces a gap in the spectrum. We shall see explicit examples of such orthonormal, discrete bases in Sec.~\ref{llmetc}. Furthermore, the CFT is a \emph{background independent} description of gravity in the interior of $AdS_d$: it is defined over the conformal boundary of the spacetime, $\mathbb{R}_t \times S^{d-2}$, where $\mathbb{R}_t$ parameterizes time. (The compactness of $S^{d-2}$ reflects the discrete nature of gravity states.) Importantly, the interior geometry does not enter the definition of the dual CFT and although the geometry of the boundary does, this is not an impediment because a quantum fluctuation altering the boundary would be suppressed by an infinite action. As a result, it becomes meaningful to compare responses to probes in various microstates and to average over them, something which would be difficult to define in gravity alone. In short, with holographic duality we gain access to the toolkit of statistical mechanics.


Consider a large black hole in AdS, which is characterized by some set of commuting charges, including a mass $M$.  In the dual field theory this black hole is represented in the canonical ensemble by a thermal density matrix:
\begin{equation}
\rho_\beta = ({\rm Tr} \exp(-\beta H))^{-1} \exp(-\beta H)
\end{equation}
The response of the black hole to a probe is computed in the field theory as a thermal correlator
\begin{equation}
\label{thermcorr}
\langle \mathcal{P} \rangle_\beta = {\rm Tr}\, \rho_\beta\mathcal{P} \,,
\end{equation}
where the operator $\mathcal{P}$ is dual to the gravity probe according to the standard AdS/CFT dictionary. An intuitive example of $\mathcal{P}$ could be
\begin{equation}
\mathcal{P} = T_{\sigma\rho}(t) \, T^\dagger_{\mu\nu}(0)\,,
\end{equation}
with $T_{\mu\nu}$ being the stress tensor, which emulates dropping a graviton from near the boundary of AdS at time 0 and detecting its return after a time $t$. However, we do not wish to limit attention to probes of the type $\mathcal{P} = \mathcal{P}_2(t) \mathcal{P}^\dagger_1(0)$, so the operator $\mathcal{P}$ should be considered general.

Consider instead the microcanonical ensemble.  From this perspective, the thermal density matrix (and hence the spacetime description as a conventional black hole) should simply be an effective, slightly coarse-grained description of most of the microstates. In the dual field theory a microstate $i$ is created by acting on the vacuum with a heavy operator $\mathcal{O}^\dagger_i$. A natural object of interest is an analogue of (\ref{thermcorr}), evaluated in a microstate:
\begin{equation}
\label{statecorr}
\langle \mathcal{P} \rangle_i = \langle 0 | \mathcal{O}_i \,\mathcal{P}\, \mathcal{O}_i^\dagger | 0 \rangle
\end{equation}
The key question is how (\ref{statecorr}) varies between microstates.

We will argue that the standard properties of black holes -- the existence of a horizon and \emph{apparent} loss of information -- are artifacts of an effective, semiclassical treatment, which averages over $e^S$ black hole microstates. The individual microstates are states in quantum gravity and may not have classical descriptions in terms of a metric. In the exceptional situation that a microstate has a description in classical gravity, the metric is necessarily horizonless because the entropy of one microstate is zero.  On the other hand coarse-grained observables should be unable to tell the microstates apart and thus will effectively see a substantial entropy.  In this way, a horizon will be seen as an artifact of our (semiclassical) inability to distinguish gravitational microstates.

To verify this picture, our strategy will be to compare expressions (\ref{thermcorr}) with (\ref{statecorr}). We expect that the thermal correlator (\ref{thermcorr}), which corresponds to the semiclassical black hole, is an ensemble average of the correlators (\ref{statecorr}) evaluated in individual microstates\footnote{Here we ignore the subtleties arising if $\mathcal{P}$ does not commute with the Hamiltonian, and assume that $\mathcal{P}$ satisfies the usual conditions for canonical ensemble treatment.}:
\begin{equation}
\label{bhisaverage}
\langle \mathcal{P} \rangle_\beta = \frac{\sum_i e^{-\beta E_i} \langle \mathcal{P} \rangle_i}{\sum_i e^{-\beta E_i}}
\end{equation}
The key in preserving information is that microstates differ from one another and their plurality has the capacity of preserving information. The differences among microstates, which encode the information fallen into a black hole, are quantified by the variance of the correlators,  $\sigma^2(\langle\mathcal{P}\rangle_i)$. A properly normalized quantity is the ratio of the standard deviation in $\langle\mathcal{P}\rangle_i$ to the mean, ${\sigma(\langle \mathcal{P} \rangle_i)}/{\langle \mathcal{P} \rangle_\beta}$.
When this is of order unity, information is preserved and readily available and conversely, when it vanishes, information is lost. Based on the argument of the previous subsection, we expect that:
\begin{equation}
\label{infloss}
\lim_{\hbar \rightarrow \infty} \frac{\sigma(\langle \mathcal{P} \rangle_i)}{\langle \mathcal{P} \rangle_\beta} = 0
\end{equation}
This is the statement that information is lost in the semiclassical limit.  A significant part of this review is devoted to studying when (\ref{infloss}) holds.

It is useful to spell out the ingredients which are necessary for the proposed resolution of the information paradox:
\begin{enumerate}
\item There must exist a set of quantum gravity states (microstates), which are responsible for the entropy of the black hole. When a CFT dual is available, these are best defined and studied in the dual field theory.
\item The behavior of the black hole must be recovered as an average over the ensemble of microstates, as in eq.~(\ref{bhisaverage}).
\item The microstates must be capable of encoding information fallen into a black hole: $\sigma^2(\langle \mathcal{P} \rangle_i) \neq 0$.
\item Information must be lost in the semiclassical limit, so we may speak of the black hole as a universal description of all microstates as they appear to semiclassical observers. This is the content of eq.~(\ref{infloss}).
\end{enumerate}



\subsection{The scale $\exp{(-S)}$: distinguishing basis microstates}
\label{eSbasis}

Point 3 above required that information be preserved and encoded in the different responses of individual microstates to probes. We now present two independent arguments, which show that in order to take advantage of the non-zero variance $\sigma^2(\langle \mathcal{P} \rangle_i) \neq 0$ to recover information from black holes, one must be able to perform measurements with a resolution of order $\exp{(-S)} = \exp({-A/4 G_N \hbar})$. 
This vanishes very rapidly in the limit $\hbar \rightarrow 0$. Consequently, in the strict semiclassical limit an observer can recover information from a black hole only if she can muster infinite precision or patience, which is in agreement with Point 4.

\paragraph{Gravity argument:} Consider an orthonormal basis of microstates, which are eigenstates of a complete set of commuting observables. In the absence of symmetries other than time translation invariance, this becomes a basis of non-degenerate Hamiltonian eigenstates.  In order to account for a degeneracy of $\mathcal{O}(e^{S})$  the energy gaps in the spectrum must be $\mathcal{O}(e^{-S})$.  Evidently, all one has to do in order to identify a basis microstate is to measure its energy (and other conserved charges).   Now, in any theory of gravity  the mass (total energy) can be measured at infinity.  Thus an asymptotic observer will be able to distinguish microstates by measuring their mass and need not experience information loss!  However, this requires access to energy resolutions of the order $\Delta E \sim \exp{(-S)}$. To see this, note that an apparatus with a resolution $\Delta E$ interacts with
\begin{equation}
\frac{d\,\Omega(E)}{dE} \Delta E \approx \frac{d\,e^{S}}{dE} \Delta E = e^S \frac{dS}{dE} \Delta E
\end{equation}
states \cite{Balasubramanian:2006iw} and thereby effectively interacting with a system which has entropy:
\begin{equation}
\mathcal{S} = S + \log \frac{dS}{dE} + \log{\Delta E}
\end{equation}
Up to logarithmic corrections, this is equal to the Bekenstein-Hawking entropy of the underlying black hole \emph{unless} $\Delta E$ can be arranged to fall exponentially with $S$. The quantity $\mathcal{S}$ will only vanish, signaling a detection of an individual microstate, if
\begin{equation}
\Delta E \sim \exp{(-S)}\,.
\end{equation}
Such a measurement would necessarily extend over a time scale exceeding the Poincar\'e recurrence time:
\begin{equation}
\Delta T \gtrsim (\Delta E)^{-1} \sim \exp S
\end{equation}
A more complete version of this argument, which applies to quantum superposition microstates, is given in \cite{Balasubramanian:2006iw}.   Thus, while for a generic black hole in AdS space all information about the microstates is present at infinity in the mass of the system thereby avoiding any fundamental information loss, a semiclassical observer with finite energy resolution simply cannot access this information.

\paragraph{CFT argument:} Suppose you drop something into a thermal black hole and wait for a time $\Delta T$ before attempting to recover it.  In the dual CFT, the thermal black hole is described as a density matrix over the accessible microstates.  However, the probe is in a pure state and we can ask what it would take to distinguish the small deviations from a thermal density matrix that must be present at late times.  In CFT, the result of the experiment we are describing is quantified by a thermal expectation value:
\begin{equation}
\label{thermunit}
F(\Delta T) \equiv \langle \mathcal{P}(\Delta T) \mathcal{P}^\dagger(0) \rangle_\beta
\end{equation}
Information is lost if $F(\Delta T)$ vanishes at late times. Thus, the scale at which information is conserved can be read off from the late time behavior of (\ref{thermunit}).  By general arguments in statistical physics, the function $F(\Delta T)$ initially decays as the contributions of different ensemble states decohere, but eventually it begins to hover around some limiting non-zero value. This happens when the phases in
\begin{equation}
\label{ft}
F(\Delta T) = \frac{\sum_{ij} e^{-\beta E_i} \langle i | \mathcal{P}(\Delta T) | j \rangle \langle j| \mathcal{P}^\dagger(0) | i \rangle}{\sum_i e^{-\beta E_i}} = \frac{\sum_{ij}e^{-\beta E_i} |\langle j | \mathcal{P}^\dagger | i\rangle|^2 e^{-i(E_j - E_i)\Delta T}}{\sum_i e^{-\beta E_i}}
\end{equation}
become randomized, that is after the interval $\Delta T$ exceeds the reciprocal of the minimal level spacing
\begin{equation}
\label{timescale}
\Delta T \gtrsim {\Delta E}/{e^S}\,,
\end{equation}
where $\Delta E$ denotes the overall energy spread among the microstates. Assuming that $\Delta E \ll \beta^{-1}$ so all the Boltzmann factors are comparable, and further, assuming that the matrix elements of $\mathcal{P}$ are of order unity, at the late times satisfying (\ref{timescale}) $F(t)$ reduces to an average of $e^{2S}$ random phases, viz. eq.~(\ref{ft}). We conclude that information is preserved on scales
\begin{equation}
\label{esscale}
\lim_{\Delta T \rightarrow \infty} |F(\Delta T)| \approx e^{-S}\,.
\end{equation}
Thus, again, unitarity of the theory is preserved in corrections of $\mathcal{O}(e^{-S})$ which cannot be accessed by semiclassical observers. 

The point that restoration of unitarity arises in AdS/CFT at $\mathcal{O}(e^{-S})$ was emphasized in \cite{Maldacena:2001kr} via consideration of the contributions of different saddlepoints of the Euclidean action (see further discussion in Sec.~\ref{addsaddle}).

\subsection{The scale $\exp{(-S)}$: distinguishing superposition microstates}
\label{eSsup}

The arguments of Sec.~\ref{eSbasis} explain why identifying a black hole microstate (decoding information) can only happen by measurement on scales of $\mathcal{O}(\exp{(-S)})$. However, both arguments implicitly assumed that the black hole was in an ensemble of Hamiltonian eigenstates. In particular, the gravity argument posits that an exact measurement of energy at infinity singles out a microstate while on the gauge theory side, eq.~(\ref{ft}) computes the magnitude of the unitarity-preserving signal from the decoherence behavior of a family of Hamiltonian eigenstates. But there is no reason to assume that a black hole must always be found in a Hamiltonian eigenstate. We shall now see that as a result of considering black hole superposition microstates, information recovery is made more difficult by an additional factor of $\exp{(-S)}$.

Consider first that a general microstate of spacetime could have a wavefunction with support on regions of configuration space with many different geometries and topologies, or even without any geometric interpretation all.   Thus in general to talk about about the microstate of a black hole spacetime we have to take a quantum cosmological perspective and think in terms of a wavefunction of the universe with appropriate boundary asymptotic boundary conditions. It is only sensible to speak of a semiclassical geometry when the wavefunction is sharply peaked \cite{Balasubramanian:2007zt} on an appropriate set of configurations.


Let us imagine a spacetime which is in a generic superposition of microstates which have the quantum numbers of a black hole.   This superposition is to be interpreted in the sense of quantum cosmology -- the universe is repeatedly prepared in an identical microstate, which is then repeatedly probed by operators $\mathcal{P}$. A superposition wavefunction is then experimentally characterized by the expected outcome of  probe experiments:
\begin{equation}
\label{ppsi}
\langle \mathcal{P} \rangle_\psi = \langle \psi | \mathcal{P} | \psi \rangle\,,
\end{equation}
where $|\psi \rangle$ is the microstate wavefuction. Thus, in quantum cosmological settings, the usefulness of a probe for recovering information is properly characterized by the variance in the expectation values (\ref{ppsi}), that is $\sigma^2 (\langle \mathcal{P}\rangle_\psi)$.

It turns out that the variance in expectation values computed over all superposition states is suppressed relative to the variance over the Hamiltonian eigenbasis \cite{Balasubramanian:2007qv}, and that the suppression factor is again $\exp{(-S)}$:
\begin{equation}
\label{2sigmas}
\sigma^2 (\langle \mathcal{P}\rangle_\psi) \approx \frac{\sigma^2 (\langle \mathcal{P}\rangle_i)}{e^S}
\end{equation}
It is easiest to see this in the case, where $\mathcal{P}$ commutes with the Hamiltonian and where the microcanonical ensemble techniques apply ($\Delta E \ll \beta^{-1}$). Expand the superposition state $|\psi\rangle$ in the Hamiltonian eigenbasis
\begin{equation}
|\psi\rangle = \sum_i^{\exp{S}} c_i |i\rangle
\end{equation}
and write down the variance in expectation values as the integral over the space of wavefunctions:
\begin{equation}
\label{sigmapsi}
\sigma^2 (\langle \mathcal{P}\rangle_\psi) = \frac{\int \mathcal{D}\Psi \langle \mathcal{P}\rangle_\psi^2}{\int \mathcal{D}\Psi} - \left( \frac{\int \mathcal{D}\Psi \langle \mathcal{P}\rangle_\psi}{\int \mathcal{D}\Psi} \right)^2 =
\sum_{ij} \langle \mathcal{P} \rangle_i \langle \mathcal{P} \rangle_j \frac{\int \mathcal{D}\Psi |c_i|^2 |c_j|^2}{\int \mathcal{D}\Psi}
- \left( \sum_{i} \langle \mathcal{P} \rangle_i \frac{\int \mathcal{D}\Psi |c_i|^2}{\int \mathcal{D}\Psi}
\right)^2
\end{equation}
The two integrals on the right hand side are independent of the detailed dynamics and depend only on the dimensionality of the Hilbert space, which is by definition $\exp{S}$:
\begin{eqnarray}
\frac{\int \mathcal{D}\Psi |c_i|^2}{\int \mathcal{D}\Psi} & = & \frac{1}{e^S} \\
\frac{\int \mathcal{D}\Psi |c_i|^2 |c_j|^2}{\int \mathcal{D}\Psi} & = & \frac{\delta_{ij}+1}{e^S(e^{S}+1)}
\end{eqnarray}
Plugging these into eq.~(\ref{sigmapsi}), we obtain
\begin{equation}
\sigma^2 (\langle \mathcal{P}\rangle_\psi) = \frac{e^S-1}{e^{2S}(e^S+1)}\sum_i \langle \mathcal{P} \rangle_i^2 - \frac{2}{e^{2S}(e^S+1)} \sum_{i<j} \langle \mathcal{P} \rangle_i \langle \mathcal{P} \rangle_j\,,
\end{equation}
which confirms eq.~(\ref{2sigmas}) since the variance in Hamiltonian eigenstates is simply:
\begin{equation}
\sigma^2 (\langle \mathcal{P}\rangle_i) = \frac{1}{e^S}\sum_i \langle \mathcal{P} \rangle_i^2 - \left( \frac{1}{e^S} \sum_i \langle \mathcal{P} \rangle_i \right)^2 = \frac{e^S-1}{e^{2S}}\sum_i \langle \mathcal{P} \rangle_i^2 - \frac{2}{e^{2S}} \sum_{i<j} \langle \mathcal{P} \rangle_i \langle \mathcal{P} \rangle_j
\end{equation}
Ref.~\cite{Balasubramanian:2007qv} presents a more complete derivation, which shows that (\ref{2sigmas}) applies so long as $\mathcal{P}$ is a finitely local observable.

The relation (\ref{2sigmas}) represents a limitation of the semiclassical observer's ability to decode the information held by a black hole that goes beyond the very high precision required to detect specific operator eigenstates as different from the thermal density matrix (Sec.~\ref{eSbasis}).   Here we see that the statistical ``de-phasing'' that  occurs in a generic state in the Hilbert space will further suppress the variance of any observable by $e^S$.


\subsection{The scale $\exp{(-S)}$: a summary}
\label{bhvsthermal}
In order for a gravitational system to be effectively described by a black hole, it must have an exponentially large degeneracy $\sim \exp{S}$ and an exponentially small level spacing $\delta E \sim \exp{(-S)}$, where $S \propto 1/\hbar$ and diverges in the semiclassical limit, precipitating semiclassical information loss. In particular, recovering information from a black hole requires an observer with access to:
\begin{itemize}
\item[a)] resolution of order $\exp{(-S)}$,
\item[b)] time scales of order $\exp{S}$,
\end{itemize}
We arrived at these observations entirely by applying {\it statistical} arguments to gravitational states.
The AdS/CFT correspondence provides a sharp definition of such states and guarantees that they may be studied using statistical mechanics. However, one may worry that the argument does not use any features unique to black holes and may therefore be applied to any thermal object in gravity, like thermal AdS space.    That is indeed correct -- it appears that, at least on the CFT side of the AdS/CFT correspondence, black holes are simply conventional thermal enembles and may be treated statistically in exactly the same way.  The only exotic thing about the black hole ensemble is the very large density of microscopic states.  This is where the dynamics of quantum gravity (or the dual strongly coupled gauge theory)   comes into the problem, by setting the nonperturbatively tiny ($\mathcal{O}(e^{-A/4 G_N \hbar})$) energy gaps that are required to account for the entropy.

But in classical gravity the {\it sine qua non} of a black hole is its horizon which certainly makes it seem very different from a thermal gas.   While the statistical arguments that we have outlined explain how semiclassical information loss can be consistent with microscopic information recovery, they do not suffice to explain why black hole entropy is proportional to a geometric quantity -- the area of the horizon.  Neither do our arguments explain how we should understand the causally disconnected black hole interior that is present in the semiclassical description or the observations of infalling observers traversing this interior region.  
The problem of understanding horizon formation and the semiclassical black hole interior in the dual CFT, though not logically necessary for resolving the information paradox, is very interesting in its own right.   It is difficult, because it is set in the realm of strongly coupled gauge theory, of which little is known quantitatively. Some material related to this question is reviewed in Sec.~\ref{microgeometries}.

\subsection{Black holes as mirrors}
\label{secmirror}

The consequences of the holographic argument for unitarity may be even more profound: it has been argued \cite{Hayden:2007cs} that if black holes are not information sinks, they must be information mirrors! Specifically, Ref.~\cite{Hayden:2007cs} posits that, as a matter of principle, any $k$ bits of information may be recovered from a black hole whose size (entropy) is $n$ bits after observing $(n+k)/2$ bits expelled from it as Hawking radiation. This implies that from an information theoretic standpoint, the lifeline of a black hole is divided into two stages:
\begin{enumerate}
\item Before the black hole has radiated away half of its information content, any additional infalling information gets temporarily trapped. 
\item The black hole evaporation process reaches its half-way point, whereafter any additional infalling and all previously infallen information becomes accessible, one bit at a time, with each bit of Hawking radiation released by the black hole.
\end{enumerate}
To wit, a black hole past its information theoretic halflife is an information mirror.

It is best to first describe the procedure by which an outside observer may recover the black hole-reflected information and then discuss the underlying assumptions. For the sake of clarity, we present the argument in its classical form. The quantum analogue, with the usual substitutions of ``bit'' with ``qubit'', ``knows'' with ``holds a system perfectly entangled with'', and ``typical permutation'' with ``a unitary transformation chosen uniformly with respect to the Haar measure'', is given in Ref.~\cite{Hayden:2007cs}.

Think of the internal state of a black hole as a classical bit string of length $n-k$, and model black hole dynamics as a typical ({\bf Assumption 3}), deterministic permutation of the $2^{n-k}$ bit strings (not of the $n-k$ bits). An outside observer has been collecting information about the black hole since its formation and now knows ({\bf Assumption 1}) both the internal state of the black hole (one of the $2^{n-k}$ strings) and the dynamical permutation (one of the $(2^{n-k})!$ such permutations). Now $k$ unknown bits are dropped into the black hole and appended to its internal state to form a new string of length $n$. The black hole dynamically processes its information content by applying one of $(2^n)!$ permutations (known to the outside observer by {Assumption 1}) \emph{before} releasing one bit of Hawking radiation ({\bf Assumption 2}). The black hole repeats the permutation-release cycle indefinitely.

In this model, the observer will decode the $k$ fallen bits of information after detecting $k+c$ quanta of Hawking radiation, with probability of false detection bounded by $2^{-c}$. To see this, note that after releasing $k+c$ quanta of information, the black hole's internal state is one of $2^{n-k-c}$ bit strings of length $n-k-c$. Because the black hole dynamical permutation is assumed typical, each of the $2^{n-k-c}$ current internal state strings is equally likely to have been produced by the dynamics of permutation from each the $2^{n-k}$ original black hole internal state strings. Thus, a false positive identification of the $k$ infallen quanta will randomly occur on a
\begin{equation}
\frac{2^{n-k-c}}{2^{n-k}} = 2^{-c}
\end{equation}
fraction of trials. Mechanically, the identification procedure involves selecting the $2^{n-k-c}$ possible precursor internal states consistent with the $k+c$ bits of detected Hawking radiation, and then weeding out those which do not agree with the $n-k$ previously known bits with which the black hole started out. That this procedure can be algorithmically carried out in times parametrically shorter than the black hole evaporation time is the content of {\bf Assumption 4}. We now comment on the four assumptions.

\paragraph{Assumption 1} Fundamentally, the argument assumes that one can identify the internal state of a black hole after recording its Hawking radiation for a sufficiently long time. This assumption agrees with the holographic intuition of Sec.~\ref{bhads}, though as seen in 
Sec.~\ref{eSbasis}-\ref{eSsup}
 the requisite measurements would of necessity probe the scale $\exp{(-S/\hbar)}$. In addition, the argument requires that the internal dynamics of black holes be knowable and computable, but this is in principle no different from invoking the deterministic nature of physical laws in other contexts. Indeed, there is no reason to doubt that black hole physics and quantum gravity are deterministic and amenable to scientific enquiry, regardless of how little we may understand of these areas at present.

Hayden and Preskill state in \cite{Hayden:2007cs} that {Assumption 1} begins to hold after the black hole has radiated away half of its original entropy. This is a consequence of a calculation by Page \cite{Page:1993df}, who computed the entanglement entropy seen by an observer with access to a subset of the degrees of freedom of a total system found in a random pure state. If the total system is divided into two parts, the one visible to the observer ($V$) and the one hidden from her ($H$), the result is that the observed entanglement entropy is generically almost equal to the maximal entropy of the smaller subsystem. In particular, an observer with access to more than half the degrees of freedom of the total system ($|H| \leq |V|$) can generically read off the hidden information from her experiments, which is the statement that the state of $H$ is perfectly entangled with a subsystem of $V$. As pointed out in \cite{Page:1993up}, this finding is directly relevant to black hole physics, in which outside observers have access to only those quanta of the system that have been radiated out by the black hole. Page's result states that the internal state of the black hole becomes identifiable after $V$ grows to contain at least half the quanta of the total system, that is after the mid-point of the black hole's evaporation.

\paragraph{Assumption 2} The second assumption is that the black hole's thermalization time $t_{\rm t}$ must not exceed the time between the emissions of two successive radiation quanta $t_{\rm r}$. The claim that all information dropped into a black hole is immediately encoded in the subsequent Hawking radiation needs this assumption so that the internal black hole dynamics may incorporate the infalling information in its computation of the next radiation quantum to be emitted. From the viewpoint of the dual CFT, one may expect that $t_{\rm t}$ and $t_{\rm r}$ are both given by $T^{-1}$, since the temperature of the black hole is the only scale in the theory.
With reference to the black hole membrane paradigm \cite{Thorne:1986iy} and the complementarity hypothesis (see Sec.~\ref{complementarity}), the authors of Ref.~\cite{Hayden:2007cs} arrive at different estimates. In either case, an upper bound on $t_{\rm t}$ is only necessary for the hypothesis that black holes are information mirrors (reflect information without delay). The statement that information is accessible after half a black hole's entropy has been radiated away holds independently of Assumption 2.


\paragraph{Assumption 3} The third assumption is that the black hole dynamics acts as a randomizer. Said differently, the effect of black hole dynamics should be typical in its class: for classical bit strings -- a random permutation, for quantum memory -- a unitary transformation chosen uniformly with respect to the Haar measure. The assumption is needed to ensure that the infalling $k$ bits (qubits) of information get uniformly jumbled up with the rest of the black hole's memory, so that any $k$ bits (qubits) of the future Hawking radiation are equally good for decoding the dropped message. If Assumption 3 failed, a black hole may expel sudden bursts of specific information as opposed to leaking it at a uniform pace in a maximally scrambled form\footnote{This is related to the view of black holes as efficient scramblers \cite{Sekino:2008he}; see  Sec.~\ref{complementarity}.}. For this reason, Assumption 3 is heuristically appealing, though it is difficult to evaluate its plausibility rigorously without a deeper understanding of quantum gravity.

\paragraph{Assumption 4} The last assumption is that the complexity of the computational problem faced by an outside observer attempting to retrieve information from a black hole does not somehow invalidate the rest of the argument. Indeed, there may be a tension between Assumption 3 and Assumption 4, because a typical permutation (or a typical unitary transformation) acting on a string of length $2^n$ takes $\mathcal{O}(2^n)$ steps to implement, a number commensurate with the Poincare recurrence time of the model black hole. Without belaboring the philosophical points, we point out that one cannot a priori exclude the possibility that the computational complexity of the decoding procedure may form an integral part of the black hole information puzzle. If so, it would be a manifestation of an exciting interplay between physics and complexity theory.

The authors of Ref.~\cite{Hayden:2007cs} attempted to resolve the tension between Assumptions 3 and 4 by conjecturing that black holes dynamics may make preferential use of special classes of transformations, which avoid a conflict with Assumption 4. A less speculative resolution of this problem would be interesting.

\section{Quantitative approaches}
\label{examples}

This section reviews several context in which it has been demonstrated quantitatively that an exponentially large density of gravity states is necessary and sufficient for resolving the information paradox.  Specifically, we examine families of heavy, exponentially dense gravity states and report dual field theory calculations, which show that their mutual differences vanish in the semiclassical limit as in eq.~(\ref{infloss}).

\subsection{Extremal black holes in $AdS_5\times S^5$}
\label{llmetc}

Black holes asymptotic to $AdS_5\times S^5$ have vanishing horizon areas if they preserve more than $1/16$ of the supersymmetries. In the following we concentrate on the half-BPS, extremal black holes.  These objects are stable and we have complete control over the microstates in gravity and in the dual field theory.  While this allows powerful, precise calculations, there is a price to be paid -- in this case the horizon coincides with the singularity and has vanishing area, because the statistical degeneracy is not large enough to produce a macroscopic horizon.   However, any small excess of energy beyond the extremal limit leads to a finite area horizon, making the 1/2-BPS black holes of AdS$_5$ an extremely useful model system.

\subsubsection{Background information}
\Label{ads5background}

The field theory dual to gravity asymptotic to $AdS_5\times S^5$ is $SU(N)$ super-Yang-Mills theory on $S^3 \times \mathbb{R}_t$. In the half-BPS sector, \cite{Lin:2004nb} developed a complete map between non-singular type IIB supergravity solutions on $AdS_5\times S^5$ and heavy field theory states.

\paragraph{Heavy states in field theory} $\mathcal{N}=4$ $SU(N)$ super-Yang-Mills theory contains three scalar fields, which are $N\times N$ matrices. In the half-BPS sector, highest weight representatives in each BPS multiplet are created by operators that are gauge invariant polynomials in the zero-modes of a single adjoint scalar field. The s-wave reduction leads to a Lagrangian describing $N$ fermions in a harmonic potential \cite{Berenstein:2004kk}. A basis of highest weight half-BPS states is specified by sets of increasing integers ${\cal
F} =\{f_1,\,f_2,\,\dots ,\,f_N\}$ related to excitation numbers of individual fermions via $E_i = \hbar\left(f_i +
\frac{1}{2}\right)$, $i=1,\dots ,N$. In the following, we shall make use of three other ways of representing 1/2-BPS basis states:
\begin{enumerate}
\item Individual fermions' excitations \emph{above the vacuum}:
\begin{equation}
r_i = f_i + 1 - i
\end{equation}
A basis state is graphically represented as a Young diagram with at most $N$ rows, whose row lengths are given by $r_i$.
\item The numbers of columns of length $j$ in the Young diagram:
\begin{eqnarray}
c_j & = & r_{N-j+1} - r_{N-j} \\
r_i & = & c_{N-i+1} + \ldots + c_N,
\end{eqnarray}
where we use the convention $r_0 = 0$.
\item The moments
\begin{equation}
M_k = \sum_{i=1}^N f_i^k\,,   ~~~~k = 1,\, \ldots,\, N.
\label{momentdef}
\end{equation}
These contain equivalent information, because one may recover the numbers $f_i$ from the set of $M_k$'s. This is done by finding the roots of the characteristic polynomial
\begin{equation}
P = \prod_{i=1}^N (x-f_i) = \sum_{p=0}^N (-1)^p \pi_p x^{N-p}\,,   ~~~~\pi_p = \sum_{i_1 < i_2 < \cdots i_p} f_{i_1} f_{i_2} \cdots f_{i_p}
\end{equation}
after rewriting its coefficients in terms of the moments using the Newton-Girard formula
\begin{equation}
p \pi_p + \sum_{k=1}^p (-1)^k M_k \pi_{p-k} = 0.
\end{equation}
A set of $N$ fermions in a harmonic well is integrable and the moments $M_k$ form the tower of commuting, integrable charges.
\end{enumerate}

\paragraph{Half-BPS supergravity solutions asymptotic to $AdS_5 \times S^5$} They have $S\mathcal{O}(4) \times S\mathcal{O}(4) \times U(1)$ symmetry, a 5-form flux and a constant dilaton, and were found in \cite{Lin:2004nb}:
\begin{equation}
ds^2 = -h^{-2} \, (dt+V_idx^i)^2 + h^2 \, (dy^2 + dx^idx^i) + R^2 \,
d\Omega_3^2 + \tilde{R}^2 \, d\tilde{\Omega}_3^2
\Label{llmmetric}
\end{equation}
The coefficients are given in terms of a function $u(x^1,x^2,y)$
\begin{equation}
R^2 = y \, \sqrt{\frac{1-u}{u}}, \quad \tilde{R}^2 = y \,
\sqrt{\frac{u}{1-u}}, \quad h^{-2} = \frac{y}{\sqrt{u(1-u)}}
\Label{llmfunctions1}
\end{equation}
and the one form $V$ is
\begin{equation}
V_i(x^1,x^2,y) = -\frac{\epsilon_{ij}}{\pi} \int_{\mathbb{R}^2}
\frac{u(v^1,v^2,0) \, (x^j-v^j) \,   \,
dv^1dv^2}{[(\vec{x}-\vec{v})^2 +y^2]^2} \, . \Label{llmfunctions2}
\end{equation}
Thus, the solution is entirely specified in terms of
$u(x^1,x^2,y)$, which in turn satisfies a harmonic equation in
$y$ and, consequently, is fully determined by its boundary condition on the $y=0$ plane:
\begin{equation}
u(r,\varphi,y) = \frac{y^2}{\pi} \int_{\mathbb{R}}
\frac{u(r',\varphi',0) \,\,  d^2\vec{r}'}{[(\vec{r}-\vec{r}')^2+y^2]^2}
\Label{llmufunction}
\end{equation}
The last equation uses polar coordinates for the $x^1x^2$ plane. The expression for the 5-form field strength is given in \cite{Lin:2004nb}.

The metrics (\ref{llmmetric}) are regular if and only if the function $u(x^1,x^2,0)$ is restricted to take the values $\pm 1/2$. Functions $u(x^1,x^2,0)$ that fall strictly between $-1/2$ and $+1/2$ anywhere on the $x^1x^2$ plane, when plugged into (\ref{llmmetric}), give rise to singular geometries. When $u(x^1,x^2,0)$ falls outside the range $[-1/2,+1/2]$, the metric (\ref{llmmetric}) develops closed timelike curves \cite{Milanesi:2005tp}.

A dictionary between these solutions and the field theory states has been established in \cite{Lin:2004nb}. The $x^1x^2$ plane at $y=0$ is identified with the single particle oscillator phase space in the dual gauge theory while the function $u(x^1,x^2,0)$ -- with a phase space distribution $W(p,q)$:
\begin{eqnarray}
(x^1,x^2) & \leftrightarrow & (p,q) \\
\label{phasespaceident}
u(x^1,x^2,0)+1 & \leftrightarrow & W(p,q)
\end{eqnarray}
The most well-known phase space distributions to be used on the right hand side are the Wigner \cite{wignerreview} and the Husimi \cite{husimi1, husimi2, OzoriodeAlmeida:1996sr} distributions, but the choice is immaterial in the semiclassical limit, because it corresponds to the operator ordering prescription on the field theory side. The semiclassical limit is implemented by sending
\begin{eqnarray}
(l_p^4 \leftrightarrow \hbar) & \rightarrow & 0 \label{lplanck} \\
N & \rightarrow & \infty ~~~~{\rm with}~(N\hbar)\sim L_{AdS}~{\rm fixed.}
\label{nsemiclassical}
\end{eqnarray}

\subsubsection{Information loss in an integrable theory}
\label{inflossinteg}

In the usual form of the black hole information paradox one
must explain the apparent loss of information, such as when a black hole has swallowed an elephant and emits thermal radiation. The half-BPS geometries asymptotic to $AdS_5 \times S^5$ present one with the opposite challenge: because the half-BPS sector is integrable, one expects that information is trivially preserved and the problem is instead to explain \emph{apparent} information loss. This was accomplished in Refs.~\cite{Balasubramanian:2005mg, Balasubramanian:2006jt}.

Recall that half-BPS states of $\mathcal{N}=4$ super-Yang-Mills are in one-to-one correspondence with Young diagrams with at most $N$ rows. We are interested in \emph{heavy} states, whose conformal dimensions (or masses, or the numbers of boxes in their Young diagrams) scale as $\mathcal{O}(N^2)$. This is because in the dual gravity this is the largest possible mass compatible with the asymptotics, so these operators create the heaviest objects in $AdS_5 \times S^5$. Indeed, the scale $\mathcal{O}(N^2)$ agrees with the masses of all known black holes in this background.

\paragraph{Almost all heavy states look alike} Ref.~\cite{vershik} used canonical ensemble techniques to show rigorously that in the limit of large number of boxes, almost all Young diagrams approach a certain limiting shape described by the so-called limit curve. The authors of \cite{Balasubramanian:2005mg} extended this analysis to studying Young diagrams with $\mathcal{O}(N^2)$ boxes and at most $N$ rows. When expressed in terms of the variables $c_j$ of Sec.~\ref{ads5background}, the partition function factorizes:
\begin{equation}
Z = \sum_{c_1,\ldots,c_N=1}^\infty q^{\sum_j j c_j} = \prod_{j=1}^N (1-q^j)^{-1}
\label{zsym}
\end{equation}
A choice of $q = \exp{(-\beta \hbar)}$ sets the expected conformal dimension. When the latter is $\mathcal{O}(N^2)$, $(\beta \hbar)$ is $\mathcal{O}(N^{-1})$ and consequently the entropy is of order $N$. This growth rate is insufficient to produce a macroscopic horizon area in the semiclassical limit, which echoes the statement that half-BPS black holes in $AdS_5 \times S^5$ are horizonless.

From eq.~(\ref{zsym}) one easily calculates the expectations of the numbers of $j$-height columns:
\begin{equation}
\label{cjaverage}
\langle c_j \rangle = (\exp{\beta j} - 1)^{-1}
\end{equation}
For $j \ll N$ the occupancies are large since $\beta \propto N^{-1}$. The values (\ref{cjaverage}) define a ``typical state'' represented with a smooth ``limit curve''. The significance of the limit curve is that the Young diagrams of almost all states lie close to it. In particular, deviations from (\ref{cjaverage}) are exponentially suppressed; for further details, see \cite{Balasubramanian:2005mg}.

\paragraph{The typical state represents a black hole} Compute the phase space distribution $W(p,q)$ of a Young diagram state and plug it into the metric (\ref{llmmetric}) as in (\ref{phasespaceident}). The resulting geometry is, up to $\mathcal{O}(\hbar)$ corrections in $W(p,q)$, smooth. However, applying the same procedure to the typical state produces a singular geometry. On the field theory side, the emergence of a singularity may be traced to the fact that a Young diagram with its discrete steps of row-length approaches and, in the semiclassical limit, is effectively being replaced with, a smooth curve (see \cite{Balasubramanian:2005mg} for details).   The loss of information in going from the precise microscopic description of a state to the coarse-grained summary represented by the limit curve translates in gravity into semiclassical information loss, and the appearance of a singularity. In summary, almost all field theory states, each of which can have a good, non-singular gravity dual (up to $\hbar$-corrections), share a common semiclassical description, whose gravity dual is singular. At least in this case the gravitational singularity is an artifact of imposing an effective, semiclassical description on quantum gravity states.

\paragraph{Integrable charges are encoded in metric multipoles} We have reviewed the argument showing that almost all half-BPS states look identical in the semiclassical limit, and argued that this leads to semiclassical information loss in the gravity theory. It is instructive to see how exactly information loss comes about. This is all the more interesting since, na\"\i vely, the integrable structure of the theory should guarantee that information is preserved.

The first step is to see where the integrable charges are encoded in gravity. A calculation carried out in \cite{Balasubramanian:2006jt} showed that they are in one-to-one correspondence with multipole moments of metric components. As an illustration, the function $u(x^1,x^2,y)$ that specifies the metric (\ref{llmmetric}) via eqs.~(\ref{llmfunctions1}) admits the multipole expansion
\begin{equation}
u(\rho,\theta,\phi) = \sum_{k=0}^\infty \left( \frac{\hbar}{\rho^2} \right)^{k+1} \!\!\Big(M_k + \big({\rm some~linear~combination~of~}M_1,\ldots,M_{k-1} \big) \Big)\, F_k(\cos^2 \theta),
\label{ucharges}
\end{equation}
where we have used spherical coordinates. Thus, the charges $M_k$ may be systematically extracted in gravity from the multipole moments of the metric. Now take the semiclassical limit (\ref{lplanck}-\ref{nsemiclassical}) of expression (\ref{ucharges}). On dimensional grounds alone, one easily sees that the charges $M_k$ generically scale as $N^{k+1}$. Writing
\begin{equation}
M_k = m_k N^{k+1} \left(1 + \mathcal{O}(N^{-1})\right)\,,
\label{masymptotic}
\end{equation}
eq.~(\ref{ucharges}) reduces in the limit $N \rightarrow \infty$ to:
\begin{equation}
u(\rho,\theta,\phi) = \sum_{k=0}^\infty \frac{m_k}{\rho^{2k+2}} F_k(\cos^2 \theta)
\label{uasymptotic}
\end{equation}
Thus, the conserved charges $M_k$ may be directly read off in gravity from the multipole expansion of the metric components.

\paragraph{Semiclassically the multipoles are universal} However, a semiclassical observer will not be able to distinguish microstates or extract information from a black hole. The reason is that $m_k$, the leading order piece of $M_k$ that alone survives the semiclassical limit, is the same for almost all states. The differences among microstates, which are responsible for preserving unitarity, are subleading in $N$ and are not accessible to semiclassical observers. Thus information is preserved in the full theory, but not in the semiclassical regime.

Heuristically, one confirms this by noting that in terms of the excitation numbers $f_i$ states differ from one another at $\mathcal{O}(1)$, which the definition (\ref{momentdef}) of $M_k$ translates into subleading, $\mathcal{O}(N^k)$ corrections. A more rigorous argument considers two regimes separately. For $k \ll N$, it is possible to rigorously show that
\begin{equation}
\lim_{N \rightarrow \infty} \sigma(m_k) / \langle m_k \rangle = 0\,,
\label{typicalitymk}
\end{equation}
which is a version of eq.~(\ref{infloss}). Meanwhile, a measurement of multipoles beyond $k \sim N^{1/4}$ requires either probing sub-Planckian scales or scales that diverge in the semiclassical limit. The reason is that the $k^{\rm th}$ multipole may be isolated from a $k^{\rm th}$ derivative of some metric component, but measuring that involves partitioning the measurement span into $k$ sub-segments. Using the relation
\begin{equation}
\frac{L_{AdS}}{l_P} \propto N^{1/4}\,,
\end{equation}
which states that even the largest scale in $AdS$ can only fit at most $\mathcal{O}(N^{1/4})$ Planck length segments, one concludes that the higher multipoles are semiclassically unobservable. For more details, consult \cite{Balasubramanian:2006jt}. Overall, a semiclassical observer interacts with an effective, singular geometry and observes information loss even though the fundamental theory is unitary and even integrable.

\subsubsection{Generalizations with less supersymmetry}

In the preceding subsection we looked at heavy states in $\mathcal{N}=4$ $SU(N)$ super-Yang-Mills theory. This is the most symmetric theory in the class of toric quiver gauge theories \cite{Kennaway:2007tq}, which have $\mathcal{N}=1$ supersymmetry in four dimensions. Toric quiver gauge theories are dual to type IIB string theory on $AdS_5 \times X^5$ \cite{Kehagias:1998gn, Klebanov:1998hh, Acharya:1998db, Morrison:1998cs}, where $X^5$ is a compact five-dimensional manifold satisfying certain stringent conditions. One may apply the methods of Sec.~\ref{inflossinteg} to study heavy half-BPS operators in a general $\mathcal{N}=1$ quiver gauge theory. This
generalization extends the previous analysis to other theories and to the $1/8$-BPS sector of $\mathcal{N}=4$ super-Yang-Mills. The relevant calculations were published in \cite{Balasubramanian:2007hu}.

The result is that certain generalizations of (\ref{typicalitymk}) apply and heavy operators in quiver gauge theories exhibit typicality, too. Consequently, we expect that the corresponding geometries give rise to singularities and information loss when treated semiclassically. However, a detailed map between field theory states and supergravity solutions such as \cite{Lin:2004nb} has not been developed, so at present it is not possible to verify this claim. In the meantime, a family of candidate black hole solutions has been written down \cite{Gauntlett:2006ns} as progress towards a map between field theory states and supergravity solutions continues \cite{Brown:2007xh}.

\subsection{Massless BTZ black hole}
\label{masslessbtz}

We now turn to black holes in $AdS_3$ -- the BTZ spacetimes \cite{Banados:1992wn}. To apply our methodology we examine cases in which a known conformal field theory is dual to an asymptotically AdS$_3$ theory of gravity.  One such example involves the worldvolume theory of $N_1$ D1-branes wrapped on $S^1$ and $N_5$ D5-branes wrapped on $S^1 \times T^4$, whose dual is $AdS_3 \times S^3 \times T^4$ with the $AdS_3$ radius scaling as
\begin{equation}
L_{AdS} \propto (N_1 N_5)^{1/4} \equiv N^{1/4}.
\end{equation}
As in the previous subsection, the semiclassical limit is $N \rightarrow \infty$. While we could have wrapped the D5-branes on $S^1 \times K3$, choosing $S^1 \times T^4$ is convenient because the field theory is simple. Specifically, it is the $\mathcal{N}=(4,4)$ supersymmetric sigma model on $(T^4)^N/S_N$, deformed by a set of marginal operators \cite{Strominger:1996sh, deBoer:1998ip, Seiberg:1999xz, Larsen:1999uk}. In the following we consider this theory at the so-called orbifold point, with all the marginal deformations turned off. (Recent papers aimed at extending the material presented in this section away from the orbifold point include \cite{Avery:2010er, Avery:2010hs}.) This is convenient, because the field theory at the orbifold point is free.  For the sector of BPS states moving away from the orbifold point of the field theory will not change counting tasks such as evaluating the degeneracy or checking whether there is a typical state.

\subsubsection{Background information}

The BTZ family of $AdS_3$ black holes \cite{Banados:1992wn} is parameterized by the mass $M \geq 0$. In the following we concentrate on the BPS, massless solution:
\begin{equation}
\label{btzmetric}
ds^2 = -\frac{r^2}{L_{AdS}^2} dt^2 + \frac{L_{AdS}^2}{r^2}dr^2 + r^2 d\phi^2
\end{equation}
According to the AdS/CFT dictionary, one can recover a CFT two-point function of an operator by considering the bulk-boundary propagator of the dual spacetime field and sending the bulk point to the boundary. For a field of conformal weights $(1, 1)$ this yields \cite{KeskiVakkuri:1998nw}:
\begin{equation}
\label{btzcorr}
\langle \mathcal{A}(w) \mathcal{A}(0) \rangle =
\sum_{k=-\infty}^{\infty} \frac{1}{(w-2\pi k)^2(\bar{w}-2\pi k)^2}
\end{equation}
In the above, we expressed the separation between the operator insertions in terms of:
\begin{equation}
w = \phi - t / L_{AdS} \qquad \bar{w} = \phi + t / L_{AdS}
\end{equation}
Eq.~(\ref{btzcorr}) represents the massless BTZ black hole in the natural parlance of the dual CFT, that is in terms of correlation functions.

The BTZ black hole has the boundary conditions and quantum numbers of a ground state in the Ramond sector of the D1-D5 CFT \cite{Strominger:1997eq}. Such ground states are built of bosonic ($\sigma^\mu_n$) and fermionic ($\tau^\mu_n$) twist operators, which create winding sectors of the worldsheet that wrap $n$ copies of the $T^4$. The superscripts $.^\mu$ correspond to global symmetries of $S^3 \times T^4$ and range from 1 to 8. Physical states are required to carry total twist $N$. Overall, the Ramond ground states are created by composite twist operators $\sigma$
\begin{equation}
\label{sigmadef}
\sigma = \prod_{n\mu} (\sigma^\mu_n)^{N_{n\mu}} (\tau^\mu_n)^{N'_{n\mu}}\,,
\end{equation}
which in turn correspond to partitions of $N$ of the following type:
\begin{equation}
\label{d1d5partition}
N = \sum_{n=1}^N \sum_{\mu = 1}^8 n(N_{n\mu} + N'_{n\mu}), \qquad N_{n\mu} = 0,1,\ldots, \qquad N'_{n\mu} = 0,1.
\end{equation}
Canonical ensemble techniques analogous to those used in Sec.~\ref{inflossinteg} reveal that in the thermodynamic limit $N \rightarrow \infty$, there are $\mathcal{O}(\sqrt{N})$ such partitions. A finite horizon area would require $L_{AdS}^4 \sim N$ (the volume of $S^3$ contributes a factor of $L_{AdS}^3$), so this is consistent with the fact that the massless BTZ black hole is horizonless.

\subsubsection{The typical state reproduces the massless BTZ black hole}
\label{microrepBTZ}

In the canonical ensemble the occupation numbers $N_{n\mu},N'_{n\mu}$ are independent and therefore their expectations are given by the usual Bose-Einstein and Fermi-Dirac values
\begin{equation}
\langle N_{n\mu} \rangle = (\exp{\beta n} -1)^{-1} \qquad \langle N'_{n\mu} \rangle = (\exp{\beta n} +1)^{-1}\,,
\label{befd}
\end{equation}
where condition (\ref{d1d5partition}) sets $\beta = \pi\sqrt{2/N}$. As in the case of $\mathcal{N}=4$ super-Yang-Mills, these values define a typical Ramond ground state, which represents the common structure of almost all Ramond ground states (\ref{sigmadef}).

Can one distinguish the states $\sigma$? For a non-twist bosonic probe of conformal weights $(1,1)$ the two-point function takes the form \cite{Balasubramanian:2005qu}
\begin{equation}
\label{d1d5corr}
\langle \mathcal{A}(w) \mathcal{A}(0) \rangle_\sigma =
\frac{1}{N} \sum_n n \sum_\mu N_{n\mu} \sum_{k=0}^{n-1} \frac{1}{[2n \sin (\frac{w-2\pi k}{2n})]^2 [2n \sin (\frac{\bar{w}-2\pi k}{2n})]^2}\,,
\end{equation}
where the numbers $N_{n\mu}$ define $\sigma$. Ref.~\cite{Balasubramanian:2007qv} analyzed the variance in this expression and showed that eq.~(\ref{infloss}) holds. Consequently, almost all states $\sigma$ respond in the same way to the probe $\mathcal{A}(w)$ in the limit of large $N$ and there is semiclassical information loss. Furthermore, at time scales $t \ll N^{1/2}$, the universal response of almost all probes reproduces the behavior of the massless BTZ black hole (\ref{btzcorr}). To see this, note that the form of $\langle N_{n\mu} \rangle$ implies that the summation over $n$ is dominated by terms of order $n \sim N^{1/2}$, but for those summands the summation over $k$ is a good approximation of eq.~(\ref{btzcorr}). Ref.~\cite{Balasubramanian:2005qu} contains a similar set of arguments for fermionic probes.

At larger times the correlator (\ref{d1d5corr}) hovers around a non-zero value, in contrast to the BTZ correlator (\ref{btzcorr}). This is in agreement with the intuition spelled out in Sec.~\ref{eSbasis}. In that section we also anticipated that the non-zero long-time average of (\ref{d1d5corr}) would be approximately $e^{-S}$, but this is not the case. One explanation is that the non-twist probes evaluated in the orbifold CFT are insufficient to explore the full set of BTZ microstates \cite{Balasubramanian:2005qu}.

\subsubsection{Spacetime microstates?}
\label{microgeometries}

The D1-D5 system has been widely used in the literature to support and illustrate the fuzzball proposal \cite{Mathur:2005zp}, which states that black hole microstates have spacetime realizations in terms of horizonless bound states of D-branes that have an extended spacetime structure.   The proposal further suggests that the transverse size of the underlying bound state is responsible for the apparent presence of a horizon, and  that the region behind this apparent horizon is in a topologically complex or perhaps even non-geometric configuration arising from the underlying bound state.  This complexity, which is not resolvable by a semiclassical observer, is then responsible for apparent information loss.   In this sense the fuzzball proposal is conceptually similar to the point of view described in Sec.~\ref{preliminaries}, although the statistical arguments presented there make no specific claims about the structure of the spacetime microstate at or behind the horizon.  Below we briefly review the ideas leading to the fuzzball proposal; for a more complete exposition see \cite{Mathur:2005zp}.

Before delving into the details, we recall a general argument in support of a spacetime realization of microstates given in \cite{Mathur:2009hf}. Its authors argue that under the assumption that the equivalence principle extends to a classical horizon it is impossible to salvage unitarity by including small corrections to Hawking radiation. If one attempts to do so, one instead finds that the entanglement between the interior and the exterior of the black hole always increases, which implies that the end result of a unitary black hole evaporation process would have to be a remnant (see Sec.~\ref{secremnants}). The best way to motivate this finding is to highlight the difference between black hole evaporation and ordinary phenomena such as burning paper \cite{Mathur:2011wg}. In the latter case, the entanglement between the exterior (photons expelled by burning paper) and the interior (the remaining, unburnt paper) is bounded by the size of either subsystem and decreases as the remainder of the paper burns up. In contrast, the process of Hawking radiation proceeds by forming entangled pairs of positive energy particles that escape to the outside and negative energy particles that fall into the interior of the black hole. These negative energy messengers increase the entanglement while decreasing the mass of the black hole, which forces the black hole to eventually form a small mass / high entropy object otherwise known as a remnant. Note that this argument directly contradicts Assumption~1 of \cite{Hayden:2007cs} (see Sec.~\ref{secmirror}), which applies to black holes the result that the entropy of a subsystem is bounded by its size \cite{Page:1993df}. In summary, Ref.~\cite{Mathur:2009hf} claims that if we exclude remnants, the only way to salvage unitarity is to introduce large corrections at the location of a putative horizon, for instance those effected by fuzzballs.

\paragraph{The Lunin-Mathur geometries}

To each Ramond ground state (\ref{sigmadef}) corresponds a spacetime geometry without horizons \cite{Lunin:2001jy}. After a U-duality, the D1-D5 system maps to the fundamental string carrying $N_1$ units of momentum and winding $N_5$ times around an $S^1$. One may then U-dualize back the metric of the fundamental string, which is known from the null chiral model \cite{Tseytlin:1996yb} for an arbitrary classical string profile ${\bf x} = {\bf F}(v)$. The resulting metric is smooth and horizonless and asymptotes to $\mathbb{R}^{4,1} \times S^1 \times T^4$ coordinatized by $({\bf x}, t, y, {\bf z})$. In the string frame it takes the form:
\begin{eqnarray}
\label{d1d5metric}
ds^2 & = & \frac{1}{\sqrt{f_1 f_5}} \Big( -(dt +A)^2 + (dy + B)^2 \Big) + \sqrt{f_1 f_5}\, d{\bf x}^2 + \sqrt{\frac{f_1}{f_5}}\,d{\bf z}^2 \nonumber \\
e^{2\Phi} & = & \frac{f_1}{f_5} \equiv \left({1 + \frac{Q_5}{L}\int_0^{L} \frac{|\dot{\bf F}(v)|^2 \,dv}{|{\bf x} - {\bf F}(v)|^2}} \right) \left/ \vphantom{\frac{Q_5}{L}} \right. \Bigg({1 + \frac{Q_5}{L}\int_0^{L} \frac{dv}{|{\bf x} - {\bf F}(v)|^2}} \Bigg) \nonumber \\
A & = & \frac{Q_5}{L} \int_0^{L} \frac{\dot{\bf F}(v) \,dv}{|{\bf x} - {\bf F}(v)|^2} \qquad \qquad dB = *_4 dA
\end{eqnarray}
The classical profile ${\bf F}(v = t - y)$ has periodicity $L \propto N_5$. For bosonic ground states it is the shape of the fundamental string to which the momentum mode $\alpha^\mu_{-n}$ was applied $N_{n\mu}$ times; when fermionic modes are turned on, the expression for ${\bf F}(v)$ becomes more complicated \cite{Taylor:2005db}. The near-horizon limit of the metric (\ref{d1d5metric}) removes the terms $1 + \ldots$ from the definitions of $f_1,f_5$ and the resulting geometries are asymptotic to $AdS_3 \times S^3 \times T^4$.

We already know that there are $\mathcal{O}(\sqrt{N})$ geometries (\ref{d1d5metric}), because they correspond to the ground states $\sigma$ counted in eq.~(\ref{d1d5partition}). It has been shown on very general grounds \cite{Strominger:1996sh, Callan:1996dv} (see \cite{Mathur:2005zp} for a more pedestrian account) that bound states of D-branes carrying three different charges, such as the D1-D5-P system, carry entropy that in the large $N$ limit scales appropriately to account for finite horizon areas of black holes.

\paragraph{The emergence of the $M=0$ BTZ black hole}

Let $r$ be the radial coordinate in the $\mathbb{R}^4$ parameterized by ${\bf x}$. For most states ${\bf F}(v)$ is so complex that $\dot{{\bf F}}(v)$ is effectively random and as a result for $r \gg |{\bf F}(v)|$ the 1-forms $A, B$ vanish.  Consequently, in that regime the metric (\ref{d1d5metric}) reduces directly to
\begin{equation}
\label{microeffgeom}
ds^2 = \frac{r^2}{L_{AdS}^2} \left( -dt^2 + dy^2 \right) + \frac{L_{AdS}^2}{r^2} \left( dr^2 + r^2 d\Omega_3^2 \right) + \frac{L_{AdS}^4}{Q_5^2} d{\bf z}^2\,,
\end{equation}
where
\begin{equation}
\label{q1q5ads}
L_{AdS}^4 \equiv Q_1 Q_5 = \frac{Q_5^2}{L} \int_0^L |\dot{{\bf F}}(v)|^2\, dv.
\end{equation}
This is the product space of $S^3 \times T^4$ and the $M=0$ BTZ black hole (\ref{btzmetric}) with $\phi = y / L_{AdS}$. A simple argument \cite{Balasubramanian:2005qu} confirms that $|{\bf F}(v)| \ll L_{AdS}$ so the regime $r \gg |{\bf F}(v)|$ extends down to scales that are parametrically smaller than the $AdS$ radius. We conclude that outside a small core region the Lunin-Mathur geometries mimic the BTZ black hole. This is a geometric dual of the conclusion of Sec.~\ref{microrepBTZ}.


\paragraph{The emergence of an apparent horizon} 
How can bound states of D-branes wrapped on compact dimensions give rise to an apparent horizon? They must develop a transverse size. A transverse size can act as an effective horizon, because it sets the radial scale at which (a) microstates become readily distinguishable and (b) incident particles get absorbed by the D-branes, which from afar looks like falling behind a horizon.

To test this conjecture let us estimate of the size of a bound state. In the fundamental string picture the transverse size of a bound state can be retrieved from the typical oscillation of the string. The typical mode has wavenumber $\sqrt{N}$ and wavelength $L / \sqrt{N}$; to estimate the amplitude one multiplies the wavelength by the mean value of $|\dot{\bf F}(v)|$. Observe that the effective length of the string is $L \propto N_5$, the mean profile slope is from eq.~(\ref{q1q5ads}) $\smash{\overline{|\dot{{\bf F}}(v)|} \propto \sqrt{Q_1 / Q_5} \propto \sqrt{N_1 / N_5}}$ and the typical wavenumber is proportional to $\sqrt{N} \sim \sqrt{N_1 N_5}$. The radial size of the bound state is therefore independent of $N_1, N_5$ and it remains so after U-dualizing back to the D1-D5 system. We could have anticipated this result from the fact that the $M=0$ BTZ black hole is a naked singularity; the microstates of a black hole with a finite horizon area should have transverse sizes that grow with the charges. Nevertheless, we can perform a non-trivial check of the claim that the size of a fuzzball matches the horizon. Looking at eq.~(\ref{d1d5metric}), the surface area of the eight-dimensional hypersurface enclosing the bound state is
\begin{equation}
\Big(f_1 f_5\Big)^{-1/4} \cdot \Big(f_1 f_5\Big)^{3/4} \cdot \left( \frac{f_1}{f_5} \right) \cdot \left( \frac{f_5}{f_1} \right) \propto (N_1 N_5)^{1/2} \sim N^{1/2}\,,
\end{equation}
where the successive terms denote volumes of the $S^1$, $S^3$ and $T^4$ and the factor of $\exp{(-2\Phi)}$ that takes us to the Einstein frame. Thus, the area enclosing the bound state reproduces the microscopic count of microstates. In fact, for some 2-charge black holes one may obtain a finite horizon by including higher derivative terms in the supergravity Lagrangian \cite{Dabholkar:2004yr, Dabholkar:2004dq}; in such cases the horizon size also matches the microscopic count. For a recent review of the fuzzball proposal, consult \cite{Chowdhury:2010ct}.

Ref.~\cite{Mathur:1997wb} argued on general grounds that the transverse size of a 3-charge bound state should match the finite horizon area of the corresponding black hole, which is known to agree with the number of string theory states \cite{Strominger:1996sh, Callan:1996dv}. There is an extensive literature on constructing candidate spacetime realizations of such microstates; early works include \cite{Bena:2006is, Balasubramanian:2006gi, Berglund:2005vb, Bena:2005va} while Sec.~5 of \cite{Balasubramanian:2008da} contains a review. Interestingly, for some of these one can explicitly see how the transverse size grows from 0 to a finite value as one increases the string coupling \cite{Denef:2002ru, Balasubramanian:2006gi}. Another interesting class of geometries, called the scaling solutions \cite{Bena:2006kb, Denef:2007vg}, has a classical moduli space, which na\"\i vely includes geometries with arbitrarily deep AdS throats. Such solutions are problematic, because in the cases with holographic duals their existence should imply that the spectrum of the CFT is continuous \cite{Bena:2007qc}. However, as shown in \cite{deBoer:2008zn}, quantization effectively caps the throats at a finite depth, because an infinite throat requires a localization in phase space that is forbidden by the uncertainty principle. This is another example of how quantum effects can cure a distressing aspect of classical geometries.

\paragraph{A geometric mechanism of information loss}

Consider a graviton falling towards the throat of a microstate geometry. In the CFT picture, it is absorbed into the bound state and creates one left and one right mover on a component string. The two excitations will collide again and possibly re-emit the graviton after a time that is $\mathcal{O}(\sqrt{N})$. On the gravity side, the graviton enters the (topologically) complex interior of a fuzzball, bounces and escapes. For certain candidate microstate geometries with extra symmetry the return time has been calculated \cite{Lunin:2001dt} and found to agree precisely with the CFT calculation, including the coefficient. In this way the fuzzball proposal seems to explain the appearance of information loss: a quantum that seems to have disappeared behind a horizon wanders about the labyrinthine interior of a fuzzball. Information is preserved because it will eventually find an exit.

\subsection{Towards dynamics -- Matrix models}

In the preceding sections we have illustrated the idea that the information paradox is an artifact of the semiclassical description of a quantum system, each of whose microstates evolves without violating unitarity. The examples we looked at involved one parameter $N$, which controlled the size of the system and the relative magnitude of quantum effects. In particular, the examples of Secs.~\ref{llmetc}-\ref{masslessbtz} were free theories at $T=0$. (Note that when we used an inverse ``effective  temperature'' $\beta$ in eqs.~(\ref{zsym}, \ref{befd}), it was only as a calculational shortcut to obtaining an effective description of heavy states.)

We now turn to calculations that track the effect of a non-zero coupling constant and temperature. 
To do this,
we look at the evolution of a disturbance in a thermal background and ask whether the disturbance remains detectable at late times. In these settings, information loss may still arise from a growing density of states, but now there is an alternative mechanism -- the dynamics. Below we present some toy models in which information loss requires the large $N$ limit \emph{and} non-trivial dynamics.

The systems of choice are matrix models. They arise frequently as subsectors of field theories with gravitational duals, but for our purposes it suffices to demand from a matrix model that it suffer information loss in the $N \rightarrow \infty$ limit. Any matrix model with this key property is a potential laboratory for studying how information is preserved and why it appears lost.

\subsubsection{Festuccia and Liu's model}
\label{flmodel}

Ref.~\cite{Festuccia:2006sa} considers matrix quantum mechanics with the action:
\begin{equation}
\label{class}
S = N\, {\rm Tr}\! \int\! dt \sum_\alpha \left( \frac{1}{2} (D_t M_\alpha)^2 - \frac{1}{2} \omega_\alpha^2 M_\alpha^2 \right) - \int dt\, V(M_\alpha; \lambda)
\end{equation}
Here $M_\alpha$ are $N \times N$ matrices, $D_t = \partial_t - i \[A, \cdot\]$ is a covariant derivative, the frequencies $\omega_\alpha$ are all positive and $V(M_\alpha; \lambda)$ is a linear combination of single trace operators determined by the 't Hooft coupling $\lambda = g_{\rm YM}^2 N$, which is kept fixed in the large $N$ limit. This class of theories is well motivated, because it contains the bosonic sector of $\mathcal{N} = 4$ super-Yang-Mills theory on $S^3$ and, as we shall see momentarily, it suffers the expected information loss in the large $N$ limit. We follow \cite{Festuccia:2006sa} and concentrate on the specific model
\begin{equation}
\label{exemplar}
S = \frac{N}{2} {\rm Tr} \int dt \Big( (D_t M_1)^2 + (D_t M_2)^2 - \omega^2 (M_1^2 + M_2^2) - \lambda\, M_1 M_2 M_1 M_2 \Big)\,,
\end{equation}
but the results presented below apply generally to the class (\ref{class}).

The regime relevant to black hole physics is where energy scales as $N^2$. This is no different from the half-BPS sector of $\mathcal{N}=4$ super-Yang-Mills considered in Sec.~\ref{llmetc}, whose Lagrangian is also of the form (\ref{class}). However, we now require that $S \propto N^2$, which can be enforced by considering systems with at least two matrices as in eq.~(\ref{exemplar}). We deviate here from Sec.~\ref{llmetc}, where entropy scaled as $N$ as a consequence of the Lagrangian containing only a single matrix with its off-diagonal entries gauged away.

A convenient object to study is the connected Wightman function
\begin{equation}
\label{wightman}
G_+ (t) = \langle \mathcal{O}(t) \mathcal{O}(0) \rangle_\beta = \frac{1}{Z} {\rm Tr} \left( e^{-\beta H} \mathcal{O}(t) \mathcal{O}(0) \right) - C\,,
\end{equation}
where $C$ subtracts the contributions of the diagonal part of $\mathcal{O}$, which we take to be a multi-trace operator with $K$ insertions of $M_\alpha$. We keep $K$ fixed in the limit of large $N$; in the models (\ref{class}) that have holographic duals, this makes $\mathcal{O}$ dual to a string probe. The quantity (\ref{wightman}) is useful because
\begin{equation}
\lim_{t \rightarrow \infty} G_+ (t) = 0
\end{equation}
is a signature of information loss. Looking at the Fourier transform $G_+(\omega)$, we expect that all its singularities are poles, of which the one closest to the real axis controls the decay rate of $G_+(t)$ at large times \cite{Festuccia:2005pi} (see also Sec.~\ref{excursions}).

Begin with the free theory $\lambda = 0$. Not surprisingly, at finite $N$ $G_+(\omega)$ takes the form \cite{Festuccia:2005pi}
\begin{equation}
G_+(\omega) = 2\pi \sum_{ij} e^{-\beta E_i}\, |\langle i | \mathcal{O}(0) | j \rangle|^2 \,\delta(\omega - E_j + E_i)
\end{equation}
and there is no information loss. However, even in the large $N$ limit $G_+(\omega)$ remains a weighted sum of delta functions \cite{Festuccia:2006sa} and the system, once disturbed, does not thermalize. Evidently this theory needs interactions to forget information.

In fact, for $\lambda > 0$ a na\"\i ve planar calculation of $G_+(\omega)$ continues to yield a discrete spectrum. The caveat is that planar perturbation theory breaks down in the limit of large time. Heuristically, a non-zero perturbation breaks the degeneracy of energy eigenstates, reducing the level spacing to some $N$-dependent value that vanishes in the large $N$ limit. When the level spacing vanishes at infinite $N$, $G(\omega)$ becomes continuous, $G_+(t)$ decays at large times and information is lost.

The models (\ref{class}) carry several lessons. As discussed throughout this paper, the semiclassical limit (which is here a large $N$ limit) is an essential ingredient in information loss, and arises from the inability for the semiclassical observer to resolve physical data such as energy gaps with a precision of $\mathcal{O}(e^{-S})$.   The matrix models of \cite{Festuccia:2006sa} require a non-zero coupling $\lambda$ to produce such tiny gaps.  Indeed  the planar perturbative calculation of $G_+(t)$ in  \cite{Festuccia:2006sa}, which reveals no information loss, is valid for $t \lesssim 1/\lambda$.  Thus, as the coupling increases and the gap in the theory decreases, information becomes ever harder to recover.


\subsubsection{Iizuka and Polchinski's model}
\label{ipmodel}

Another matrix model demonstrates the importance of temperature \cite{Iizuka:2008hg} in information loss. Consider the following Hamiltonian:
\begin{equation}
\label{polchinski}
H = \frac{1}{2} {\rm Tr}\, \Pi^2 + \frac{m^2}{2} {\rm Tr} X^2 + M (a_i^\dagger a_i + \bar{a}_i^\dagger \bar{a}_i) + g (a_i^\dagger X a_i + \bar{a}_i^\dagger X^{\rm T} \bar{a}_i)
\end{equation}
Here $X_{ij}$ is a Hermitian matrix with canonical conjugate momentum $\Pi_{ij}$ and $a^\dagger_i,a_i$ ($\bar{a}^\dagger_i,\bar{a}_i$) are creation and annihilation operators for a complex vector particle $\phi$ (and its conjugate $\phi^\dagger$). The field $X_{ij}$ transforms in the adjoint and $\phi$ ($\phi^\dagger$) in the fundamental (antifundamental) representation of $SU(N)$, so the indices $i,j$ range from 1 to $N$. The Hamiltonian commutes with the number operators $N_\phi = a_i^\dagger a_i$ and $N_{\phi^\dagger} = \bar{a}_i^\dagger \bar{a}_i$, so the spectrum decomposes into independent sectors of definite
$(N_\phi,N_{\phi^\dagger})$. While the model (\ref{polchinski}) has no ground state, each $(N_\phi,N_{\phi^\dagger})$ sector does. In the following we restrict attention to the $(0,0)$ sector. One can accomplish that without modifying the dynamics by deforming the Hamiltonian with a quadratic term
\begin{equation}
H' = H + c (N_\phi+N_{\phi^\dagger}) (N_\phi+N_{\phi^\dagger}-1)
\end{equation}
with a sufficiently high pre-factor $c$.

Several considerations motivate (\ref{polchinski}). In comparison with the theory (\ref{exemplar}), it substitutes a trilinear fundamental-adjoint-fundamental interaction for a quartic interaction of adjoints. Thus, in terms of the double-line notation for Feynman diagrams, eq.~(\ref{polchinski}) effectively halves the basic interaction of (\ref{exemplar}), which leads to computational simplifications. The model is also closely related to a previously studied description of the fundamental string stretched between a probe D0-brane and a D0-brane black hole \cite{Iizuka:2001cw}. In addition, further studies of information loss in (\ref{polchinski}) give hints of an emergent bulk description \cite{Iizuka:2008eb}, though we shall not pursue this here. Last but not least, it suffers information loss in the limit of large $N$, so it is a useful setting for examining resolutions of the information paradox. Interestingly, as we will see, information loss occurs in the model (\ref{polchinski}) only at large temperature. 
We concentrate on the observable
\begin{equation}
e^{iM(t-t')} \langle\, {\rm T}\, a_i(t) a^\dagger_j(t) \,\rangle_T \equiv \delta_{ij} G(T, t-t')\,.
\end{equation}
The Schwinger-Dyson equation for the Fourier transform $\tilde{G}(T,\omega)$ reads:
\begin{equation}
\label{sde}
\tilde{G}(T, \omega) = \tilde{G}_0(\omega) - g^2 N \tilde{G}_0(\omega) \tilde{G}(T,\omega) \int_{-\infty}^\infty \frac{d\omega'}{2\pi} \tilde{G}(T,\omega') \tilde{K}_0 (T, \omega-\omega')
\end{equation}
The above expression involves the free $\phi$-propagator
\begin{equation}
\tilde{G}_0(\omega) = \frac{i}{\omega + i\epsilon}\,,
\end{equation}
because in the $(N_\phi,N_{\phi^\dagger})=(0,0)$ sector the thermal background does not affect the zeroth order evolution of $\phi$. In contrast, the propagation of $X$ is given by the free thermal propagator
\begin{equation}
\tilde{K}_0(T,\omega) = \frac{i}{1-e^{-m/T}} \left( \frac{i}{\omega^2 - m^2 + i\epsilon} - \frac{e^{-m/T}}{\omega^2 - m^2 - i\epsilon} \right)\,.
\end{equation}
The time ordering in the definition of $G(T,t)$ guarantees that $\tilde{G}(T,\omega)$ is non-singular in the upper half-plane. The fact that $g$ has mass dimension $3/2$ suggests that at sufficiently high frequency $\tilde{G}(T,\omega)$ reduces to its free form $1/\omega$. These observations allow us to close the integral in the upper half plane and take the residue. The resulting equation reads:
\begin{equation}
\label{sde2}
\tilde{G}(T,\omega) = \frac{i}{\omega} \left( 1 - \frac{g^2 N \tilde{G}(T,\omega)}{2 m(1 - e^{-m/T})} \left( \tilde{G}(T,\omega - m) + e^{-m/T} \tilde{G}(T,\omega + m)\right) \right)
\end{equation}
At $T=0$, this reduces to
\begin{equation}
\tilde{G}(\omega) = \frac{i}{\omega} \left( 1 - \frac{g^2 N}{2m} \tilde{G}(\omega) \tilde{G}(\omega - m)\right)\,,
\end{equation}
which enjoys a closed form solution in terms of Bessel functions:
\begin{equation}
\label{zerotemp}
\tilde{G}(\omega) = i\, \sqrt{\frac{2\,m}{g^2 N}} \, \frac{J_{-\omega/m}\left(\sqrt{\frac{2g^2 N}{m^3}}\right)} {J_{-1-\omega/m}\left(\sqrt{\frac{2g^2 N}{m^3}}\right)}
\end{equation}
In particular, the zero temperature solution (\ref{zerotemp}) has a set of poles on the real axis, the correlator is quasi-periodic and there is no loss of information. As temperature is increased, however, one may show that the solutions of (\ref{sde2}) have no poles and instead gradually take on a smooth form that signals information loss. The authors of \cite{Iizuka:2008hg} studied numerically how the solutions of (\ref{sde2}) vary with temperature and showed that at sufficiently high temperature $G(T,t)$ decays exponentially.

In summary, the model (\ref{polchinski}) undergoes information loss only when the large $N$ limit is accompanied by a sufficiently large temperature. In the examples we reviewed earlier large $N$ was responsible for the smallness of the gap and, by extension, for information loss. Temperature, on the other hand, controls how uniformly different states are populated in the canonical ensemble. The fact that in the present model information loss requires both large $N$ and a finite temperature is intuitive: in order to lose information, one needs many closely spaced states, all of which participate in the dynamics. 

\section{Additional  approaches}
\label{otherappr}

In this section we review several alternative approaches to the black hole information paradox and contrast them with the view presented in Sec.~\ref{examples}.

\subsection{Backreaction of Hawking radiation}

Ref.~\cite{Parikh:1999mf} highlighted the fact that quanta of Hawking radiation are produced in tunneling events that create pairs of particles astride the horizon, and speculated that the detailed sequence of such tunneling events could carry information out of a black hole. This possibility was recently revisited in \cite{Zhang:2009td}, which argued that there are enough tunneling sequences to account for the full entropy content of a black hole. A series of tunneling events, whose precise sequence is decided by the black hole microstate, is reminiscent of the scenarios discussed above. The difference lies in whether the horizon is to be treated as an emergent, semiclassical concept, as opposed to a bona fide spacetime site where Hawking quanta are produced. Refs.~\cite{Mathur:2009hf, Mathur:2011wg} discussed in Sec.~\ref{microgeometries} repudiate the latter possibility.

\subsection{Complementarity}
\label{complementarity}

The principle of black hole complementarity posits that black hole physics enjoys two complementary descriptions. On the one hand, no exotic phenomena alert a freely falling observer when she crosses a horizon and falls inside a black hole. On the other hand, the physics seen by static observers is well described by phenomena restricted to the exterior region.  From the latter perspective, the effective dynamics of matter near the black hole can be summarized in terms of  physical degrees of freedom on a {\it stretched horizon} -- a surface first studied in the membrane paradigm of black holes \cite{Thorne:1986iy}, defined to extend one Planck unit outward from the event horizon. The principle of complementarity states that these two descriptions of black hole physics are equivalent and cannot both be simultaneoursly accessible to any single observer -- they are in this sense complementary.  The principle further states that any attempt to confront the two perspectives will necessarily fail, or else involve physics beyond the Planck scale.

The authors of \cite{Susskind:1993mu} considered various thought experiments in which observers attempt to invalidate black hole complementarity by:  (1) attempting to establish the non-existence of a physical membrane at the stretched horizon for either static and collapse geometries,  or (2) first sampling Hawking radiation and then diving into the black hole in an attempt to duplicate information.   In every case, the observer encounters super-Planckian physics somewhere in the process.  In this way, \cite{Susskind:1993mu} argued that despite the small spacetime curvatures encountered at the horizon, a resolution of the information paradox necessarily requires control of short distance physics, i.e. a complete quantum theory of gravity. Indeed, in the controlled context of two-dimensional dilaton gravity \cite{Callan:1992rs} in which black hole complementarity was originally proposed \cite{Susskind:1993if}, the authors of \cite{Balasubramanian:1995sm} showed that the outgoing late-time Hawking radiation and low-energy infalling observers cannot simultaneously be described in a low energy effective theory in which all interactions must have center-of-mass energies below a specified cutoff.  In this way, a low-energy effective theory can either describe the asymptotic observer and the Hawking radiation flux, or the infalling observer at the horizon, even though there are no large curvatures in the problem until the observer reaches the singularity. This precise form of complementarity, and the conclusions of \cite{Susskind:1993mu}, follow in the end from the enormous relative boost of infalling and outgoing observers near a black hole horizon. 


The principle of black hole complementarity manifestly preserves unitarity; indeed, it posits that information is present and in principle accessible on or outside the black hole horizon at all times. Although the stretched horizon looks like a hot membrane in a thermodynamic description, on a fundamental level it supports quantum gravitational degrees of freedom, whose dynamics in principle determines the evolution of the system. In particular, it determines the detailed structure of the emitted Hawking radiation, which encodes the information about the initial state. In this sense, the resolution of the black hole information paradox offered by complementarity is not fundamentally different from the statistical view presented in previous sections.\footnote{But see \cite{Bousso:2010yn}, which suggests a qualitatively different resolution related to complementarity, which is supposed to arise as a by-product of any robust solution of the cosmological measure problem.} Both views contend that information loss is an artifact of imposing a semiclassical description on an essentially quantum gravitational system, and that departures of Hawking radiation from thermality preserve unitarity.  Black hole complementarity makes an additional claim: that the black hole interior enjoys a complementary description in terms of degrees of freedom located outside the horizon, and further, that no observer can combine or confront the two descriptions without relying on super-Planckian physics. This extra claim is not essential to the information paradox, however, and to date there is no precise quantitative realization of black hole complementarity that shows how the physics seen by an infalling observer may related to measurements by an asymptotic observer.

It has proved difficult over the years to refute the idea of complementarity. Ref.~\cite{Sekino:2008he} discusses one narrow miss based on the results of \cite{Hayden:2007cs} (see also Sec.~\ref{secmirror}). It shows that complementarity gets as close as possible to violating the no-cloning theorem \cite{Wootters:1982zz} (essentially the linearity of quantum mechanics) without actually violating it. One should stress, however, that such studies do not provide positive evidence for black hole complementarity. The latter would presumably have to arise from physics beyond the Planck scale.

\subsection{Nonlocality}

Complementarity states that the degrees of freedom inside a black hole are duplicated outside. In particular, a local field operator from a black hole interior must not commute with its outside-horizon duplicate. Thus, complementarity stipulates that a broad family of pairs of spacelike separated operators must not commute and ergo, requires a large scale violation of locality. Potentially, this is most embarrassing on so-called `nice slices' \cite{Lowe:1995pu, Lowe:1995ac}, which form a foliation of spacetime that interpolates smoothly between free fall frame on or inside the horizon and the fiducial observer's frame far away from the black hole. In principle, they contain information about both the outgoing Hawking radiation and any infalling matter (information), and accomplish that without stumbling upon large curvature anywhere except near the singularity. Furthermore, the four-momenta of infalling matter and outgoing Hawking radiation are small when projected onto a nice slice. Consequently, one expects that the physics on nice slices should be aptly described by effective field theory with no violations of locality.

As it turns out, effective field theory on nice slices must be supplemented by nonlocal terms. They arise whenever one writes down the full quantum gravity theory on nice slices and then truncates to low energies; the argument of the previous paragraph implicitly performed these operations in the reverse order. The presence of nonlocality on nice slices was shown in \cite{Lowe:1995pu} using the perturbative S-matrix of string theory and in \cite{Lowe:1995ac} using commutators of string field theory. Remarkably, the freely falling observer does not detect violations of locality until she is very close to the singularity \cite{Lowe:2006xm}, consistent with the reasoning that motivated the complementarity hypothesis.

The quest to exonerate complementarity from charges of unphysical nonlocality has led to a separate line of research, which examines the limits on locality in black hole physics independent of complementarity or string theory. Refs.~\cite{Giddings:2004ud, Giddings:2006sj} explored the consequences of the observation that the notion of spacelike separation can at best be approximate in a dynamical theory of gravity, especially in the presence of large blueshifts. In particular, the regions of nice slices that intersect the black hole interior and Hawking radiation are necessarily separated by large blueshifts. Consequently, nice slices are unlikely to host a local field theory \cite{Giddings:2006be}, which is a possible loophole in the standard argument for information loss. Interestingly, complementarity seems to demand a stronger violation of locality than does unitarity alone \cite{Giddings:2009ae}. Thus, complementarity appears to assume enough structure to accommodate two independent, plausible resolutions of the information paradox, a dynamical one based on nonlocality and the statistical one that is the subject of this review. For a brief summary of the locality bound and its relevance to black holes, the reader is referred to \cite{Giddings:2007pj}.

\subsection{Excursions beyond the horizon}
\label{excursions}

\begin{figure}[ht]
\begin{center}
\includegraphics[scale=0.6]{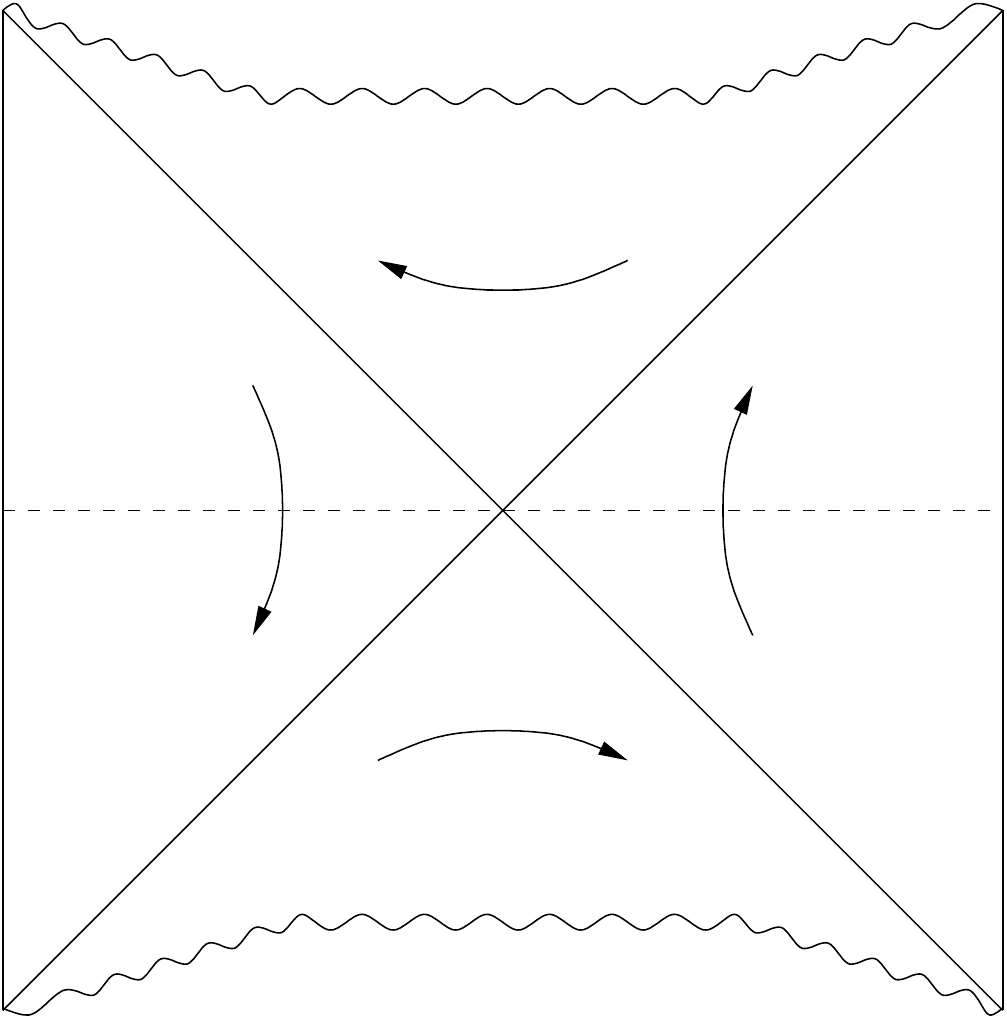}
\caption{The Penrose diagram of Schwarzschild-AdS in $d > 3$. In $d=3$, the diagram is a perfect square. Arrows mark the directions of Schwarzschild time $t$ in each region. The dashed line is fixed under the reflection symmetry $t \leftrightarrow -t$.}\label{schwarzschildads}
\end{center}
\end{figure}

The issue of information loss arises fundamentally from the existence of a spacetime horizon, and a geometric region behind it, in the semiclassical theory.  Every proposed solution to the information paradox provides some mechanism for the asymptotic observer to have access to physics from behind the horizon, e.g., because it is ``complementary" to the exterior physics, or because it is encoded in subtle Hawking radiation correlations, or because physics is nonlocal near a horizon, or because the causal disconnection of the interior is an artifact of the semiclassical limit etc.   This question becomes particularly interesting in the context of the description of eternal black holes in the AdS/CFT correspondence, because the unitary CFT is by definition supposed to encode the physics of the entire spacetime including any causally disconnected regions that might lie behind a horizon.

The extended Penrose diagram of the Schwarzschild-AdS geometry \cite{Fidkowski:2003nf} is presented in Fig.~\ref{schwarzschildads}. It contains two asymptotically AdS regions and standard holographic reasoning posits that each of them contains a conformal field theory on its boundary. The fact that we have two independent copies of the same field theory is reminiscent of the thermofield formalism, which is a way of studying thermal field theory in real time. The idea is that a particular pure state $| \Psi \rangle$ in $\mathcal{H} \otimes \mathcal{H}$, where $\mathcal{H}$ is the Hilbert space of one copy of the field theory, is capable of encoding the thermal information in the entanglement between the two $\mathcal{H}$s. It is easy to see that the correct choice of state is \cite{Balasubramanian:1998de, Maldacena:2001kr}
\begin{equation}
| \Psi \rangle = \frac{1}{\sqrt{Z(\beta)}} \sum_n e^{-\beta E_n / 2} |n\rangle_1 \otimes |n\rangle_2, \label{hhstate}
\end{equation}
because thermal expectation values of any field theory operator $\mathcal{O}_1$ living in CFT$_1$ can be obtained as expectations in $| \Psi \rangle$, the usual Boltzmann factor recovered from tracing over the states of CFT$_2$:
\begin{equation}
\langle \mathcal{O}_1 \rangle_\beta = \frac{1}{Z(\beta)} \sum_n e^{-\beta E_n}\!\! \phantom{.}_1\langle n | \mathcal{O}_1 | n \rangle_1 = \langle \Psi | \mathcal{O}_1 | \Psi \rangle
\end{equation}
Standard AdS/CFT reasoning shows that correlation functions of operators inserted on the same copy of the CFT probe the regions outside horizons -- the left and right wedges of Fig.~\ref{schwarzschildads}. Ref.~\cite{Maldacena:2001kr} proposed that that the full tensor product of the two CFTs contains information about the full extended Penrose diagram, including the regions adjacent to the singularities. The latter are probed by \emph{two-sided} correlation functions,
\begin{equation}
\langle \Psi | \mathcal{O}_1 \mathcal{O}'_2 | \Psi \rangle\,,
\label{excursioncorr}
\end{equation}
which combine operator insertions in \emph{distinct} CFTs. These are simply a technical tool in the thermofield formalism, but in the holographic context of the Schwarzschild-AdS geometry they are recovered from spacelike geodesics connecting the two boundaries. Such geodesics necessarily cross horizons.

We now show that extracting information from behind horizons requires a study of analytically continued correlation functions. The $d$-dimensional Schwarzschild-AdS geometry is given by
\begin{eqnarray}
ds^2 & = & -f(r) dt^2 + \frac{dr^2}{f(r)} + r^2 d\Omega_{d-2}^2 \label{schwadscoord}
\\
f(r) & = & \frac{r^2}{L_{AdS}^2} + 1 - \frac{r_+^2}{r^2} \left( \frac{r_+^2}{L_{AdS}^2} + 1 \right)\,,
\end{eqnarray}
where $r_+$ is the horizon radius. The Lorentzian metric (\ref{schwadscoord}) is valid in one region outside the horizon and one region inside the horizon at a time, with the boundary located at $r=\infty$ and the singularity at $r=0$. However, it can be extended to the full Penrose diagram of Fig.~\ref{schwarzschildads} by complexifying the Schwarzschild time $t$ \cite{Hemming:2002kd, Fidkowski:2003nf}.  Indeed, the Schwarzschild time in which a radial null geodesic reaches the singularity picks up an imaginary contribution from the pole on the horizon:
\begin{equation}
t = \int_\infty^0 \frac{dr}{f(r)} = -\frac{i\beta}{4} + {\rm real} \label{imaginarytime}
\end{equation}
Consequently, it is consistent to think of (\ref{schwadscoord}) as a global metric of Fig.~\ref{schwarzschildads}, with the proviso that each of the four regions has a fixed imaginary part. From eq.~(\ref{imaginarytime}) we read off that the correct assignment is to pick up an additional $-i\beta/4$ each time one crosses a horizon in the clockwise direction in the sense of Fig.~\ref{schwarzschildads}. As a matter of convention $t$ is real in the region adjacent to the right boundary. In order to maintain (\ref{imaginarytime}) for geodesics departing the left asymptotic boundary, the real part of the Schwarzschild time runs backwards in the left wedge. Overall, in each region the real part of $t$ increases in the clockwise direction, as indicated in the figure. The diagram is $t \leftrightarrow -t$ symmetric as a consequence of the $\beta$-periodicity in the imaginary time direction.

These observations define a recipe for computing the two-sided correlation functions (\ref{excursioncorr}) that contain behind-the-horizon information. All one has to do is to analytically continue the insertions from CFT$_2$ according to $t \rightarrow -t - i\beta/2$. This prescription, in principle straightforward, turns out to be tricky, because one must avoid singularities in the complex $t$-plane that arise whenever the operator insertions become lightlike separated. For BTZ black holes the correct procedure was carried out in \cite{Kraus:2002iv}. Depending on the chosen time slicing, the resulting, analytically continued correlators admit two interpretations. According to the first, one integrates interaction vertices over the entire, extended diagram of Fig.~\ref{schwarzschildads}; in the other, interactions are construed to happen only in the left and right wedges outside horizons, but the contour of the analytically continued path integral contains an additional piece imposing correlated boundary conditions on the horizons, which mimic the effect of (\ref{hhstate}). This dual interpretation is reminiscent of the ideas reviewed in Sec.~\ref{complementarity}. In the following we do not pursue the BTZ story and instead follow \cite{Fidkowski:2003nf}, who studied Schwarzschild-AdS black holes in $d>3$. These black holes are more instructive, because they contain curvature singularities and because they illustrate a further subtlety in extracting behind-the-horizon information, reviewed below.

Consider a highly massive probe in AdS space that is dual to some scalar operator $\mathcal{O}$ in the CFT.    The AdS path integral for the spacelike two-point correlator of  $\mathcal{O}$ will be dominated by geodesic paths in the bulk, so that $\langle \mathcal{O}(x_1) \mathcal{O}(x_2) \rangle \approx \exp{(-m \mathcal{L})}$ where $\mathcal{L}$ is the length of a  geodesic joining the two insertions on the AdS boundary \cite{Balasubramanian:1999zv}.  This expression requires regulation to remove the divergence in length near the AdS boundary, and some care is necessary when there are multiple geodesics especially in the case of non-static spacetime \cite{Louko:2000tp, Festuccia:2005pi}.  Neglecting the latter subtlety for the moment,  the two-sided correlator between operators inserted in the CFTs on the two boundaries of the eternal AdS black hole is given by $\exp{(-m \mathcal{L})}$, where $\mathcal{L}$ is the length of a spacelike geodesic joining the two insertions. Thus, a study of correlation functions reduces to understanding spacelike geodesics in the black hole background. To simplify the problem, \cite{Fidkowski:2003nf} noted that a general two-sided correlator, $\langle \phi_2(t_2) \phi_1(t_1) \rangle$ (where
$t_1, t_2$ are Lorentzian times in CFT$_1$ and CFT$_2$), is related to
\begin{equation}
\left\langle\!\! \Psi\! \left| \phi_2\!\left(\frac{t_1-t_2}{2}\right)\! \phi_1\!\left(\!-\frac{t_1-t_2}{2}\right)\! \right| \!\Psi\!\! \right\rangle \equiv \langle \Psi | \phi_2(-t_0) \phi_1(t_0) | \Psi \rangle \equiv \langle \phi(t_0-i\beta/2) \phi(t_0) \rangle_\beta \equiv
\langle \phi \phi \rangle (t_0)
\end{equation}
by time translation since time runs backwards in CFT$_2$. The geodesics determining $\langle \phi\phi \rangle(t_0)$ are the simplest to study due to their symmetry, which guarantees that their closest approach to the black hole singularity falls at $t = -i\beta/4$. In the following we concentrate on those and adopt a notation, in which correlators depend on a single parameter, $t_0$, which in the bulk perspective is the time at which a symmetric geodesic reaches a boundary.
  
A simple study of the geodesics shows that there exists a time $t_c<0$ such that correlation functions of opetators inserted before $t_c$ or after $-t_c$ vanish:
\begin{equation}
t_0 \not\in \[t_c, -t_c\] \quad \Rightarrow \quad \langle \phi \phi \rangle (t_0) = 0
\end{equation}
In the parlance of gravity, if a spacelike geodesic departs the boundary outside this interval, it cannot be symmetric and spacelike. In between $t_c$ and $-t_c$, however, the geodesics penetrate some finite distance past the horizon, then reverse direction and continue to the other boundary. Heuristically, one could think of such geodesics as determining the amplitudes for two virtual particles created on the different asymptotic boundaries to annihilate behind the horizon.

\begin{figure}[ht]
\begin{center}
\includegraphics[scale=0.6]{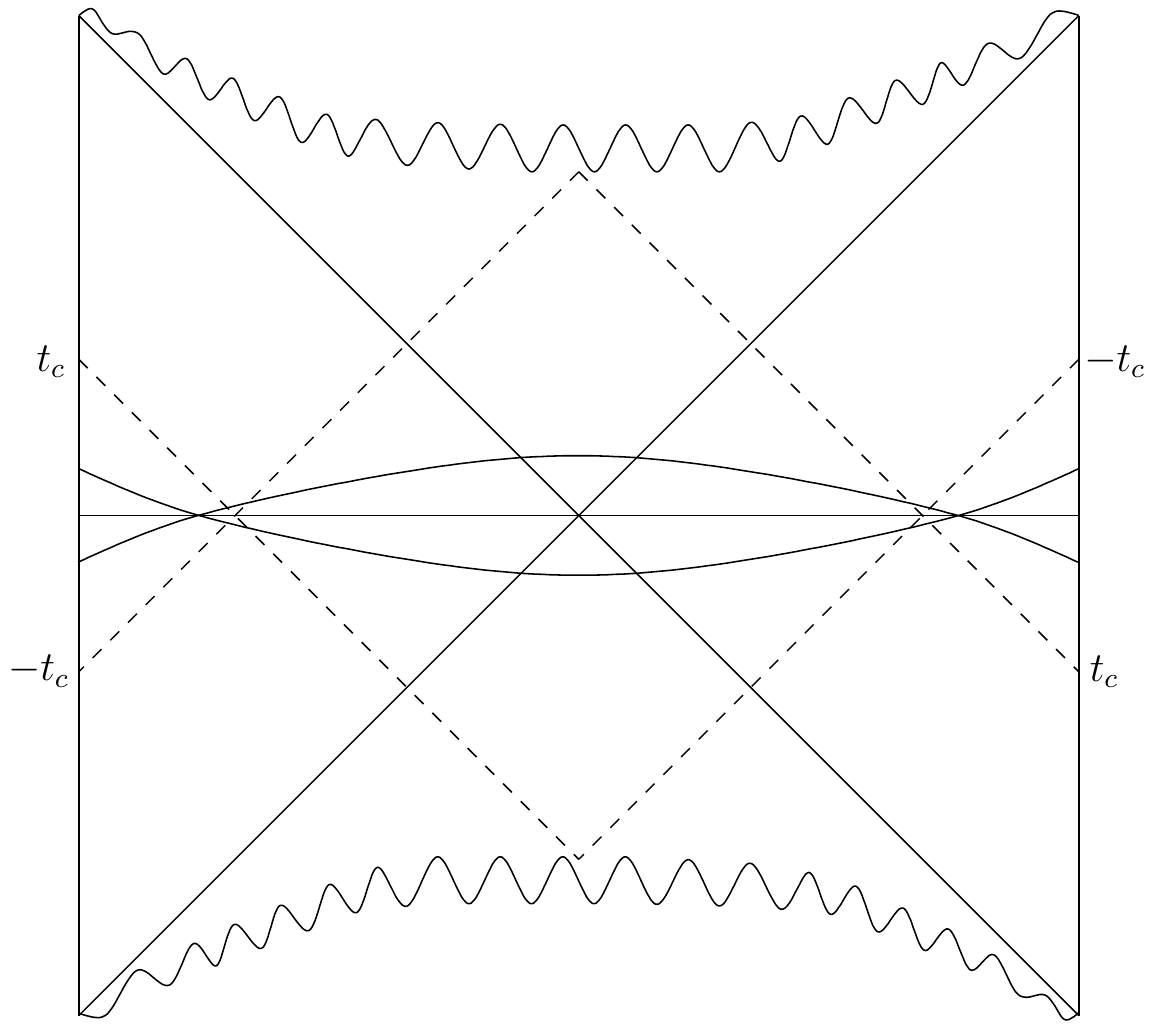}
\caption{Spacelike geodesics in Schwarzschild-AdS. For initial times $t_0 \in \[t_c, -t_c\]$ symmetric geodesics cross the horizon, reverse direction, and escape to the other asymptotic boundary; outside this interval geodesics cannot be symmetric and spacelike. Points in the bulk at $t=0$ are traversed by three distinct spacelike geodesics, precursors of the three sheets of $\mathcal{L}(t_0)$.}
\label{geodesics}
\end{center}
\end{figure}

What is the signature of the black hole singularity in this language? We know we must insert operators on two different boundaries, but within this family of correlators, some will be more useful for probing the black hole singularity than others. To wit, the $t=0$ correlator is given by the $t=0$ radial line, which skirts the horizon at a point and does not approach the singularity. The correlators with $t \sim t_c$ are more promising, because the relevant geodesics get arbitrarily close to the singularity. They also become almost null, so one may anticipate that the signature of the black hole singularity will be a pole in the correlator at $t = t_c$. The bad news for an observer hoping to peek behind the horizon is that such a pole is forbidden:
\begin{eqnarray}
\langle \phi \phi \rangle(t_0) & = & \sum_{nm} e^{-\beta(E_n+E_m)/2 - 2it(E_n -E_m)} |\langle n | \phi(0) | m \rangle|^2 \nonumber \\ & \leq & \sum_{nm} e^{-\beta(E_n+E_m)/2} |\langle n | \phi(0) | m \rangle|^2 = \langle \phi \phi \rangle(0)
\end{eqnarray}
The quantity $\langle \phi \phi \rangle(0)$ is finite and explicitly computable from the $t_0=0$ geodesic, which is simply the dashed line in Fig.~\ref{schwarzschildads}. Thus, the expected lightcone pole is absent from the two-sided Lorentzian correlator.

To understand the correlator as $t_0$ approaches $t_c$, \cite{Fidkowski:2003nf} compute the action $\mathcal{L}$ of a spacelike geodesic. Near $t_0=0$ its dependence on the boundary insertion time $t_0$ is
\begin{equation}
\mathcal{L}(t_0) \sim -t_0^{4/3},
\end{equation}
which signals that in complexified Schwarzschild-AdS three different geodesics connect a generic pair of points on the two boundaries. A way to visualize this is to consider geodesics connecting points in the bulks of the two outside-the-horizon regions. For example, for a given $s$, the boundary-boundary geodesics with $t_0=-s,0,s$ coalesce at two different spacetime points, as illustrated in Fig.~\ref{geodesics}. In the complexified Schwarzschild-AdS the triumvirate of geodesics persists across the bulk, with the exception of some non-generic points such as $(r,t)=(\infty, 0)$, where the trio becomes degenerate. The conclusion is that $\mathcal{L}(t_0)$ is a three-sheeted
Riemann surface. One of the sheets yields real geodesics in Lorentzian signature; the na\"\i vely anticipated lightcone pole at $t_0 = t_c$ in the correlator would arise from this real sheet. The other two sheets define complexified geodesics in complexified Schwarzschild-AdS.

The authors of \cite{Fidkowski:2003nf} find that in Euclidean signature for $t_0$ not too close to 0 the contributions of the complexified geodesics dominate the contribution from the real sheet. Because the Lorentzian correlators are defined as analytic continuations of their Euclidean counterparts, we conclude that the real sheet does not contribute to the real time correlator, which explains the absence of the lightcone pole. However, near $t=0$ the real sheet is dominant. Consequently, an analytic continuation of the correlator from the neighborhood of $t=0$ allows one to recover the real sheet along with the $t=t_c$ pole, a signature of the black hole singularity, despite its absence from the Lorentzian correlator. Analytic continuation is a ticket for an excursion beyond the horizon.

These findings can be used in a variety of ways. Ref.~\cite{Fidkowski:2003nf} includes a discussion of finite $m$, $\alpha'$, and $g_s$ corrections while the authors of \cite{Festuccia:2005pi} study analytically continued correlators in Fourier space. The authors of \cite{Levi:2003cx, Balasubramanian:2004zu} used similar techniques to explore the representation of the {\it inner} horizons of rotating black holes, and their instability to collapse, in the dual field theory description  From the perspective of the black hole information paradox, a salient feature of these efforts is the treatment of the horizon as being ``really there'' in spacetime, at least in the description of thermal density matrices in the field theory, and not as an artifact of an effective description of many underlying microstates.

Ref.~\cite{Fidkowski:2003nf} found that the scale $\exp{(-S)}$ is not relevant the discussion of excursions beyond the horizon.  Meanwhile \cite{Festuccia:2005pi} pointed out that the analytic continuation does not commute with the large $N$  limit.  In the strict large $N$ limit, where the semiclassical description with a horizon is valid, analytic continuation seems to permit us to probe behind the horizon.   But this analytic continuation need not have much to do with physics at finite $N$ where, as we have argued, the horizon simply arises in the course of an approximate description of  individual microstates.  Indeed, the analytic continuation to imaginary time of correlation functions computed in particular microstates will generally disagree markedly from the analytic continuation of correlation functions computed in the strictly thermal density matrix at large N \cite{Balasubramanian:2007qv}.  This suggests that studies of excursions behind the horizon, while certainly informative about the structure of physics in the semiclassical limit and the dual CFT representation of spacetime singularities, will not tell us about the recovery of information from black holes.  

\subsection{Additional saddle points}
\label{addsaddle}
In the preceding subsection we employed the Hartle-Hawking state (\ref{hhstate}) for computing correlation functions in the Schwarzschild-AdS background.  In its original derivation \cite{Hartle:1983ai}, one glues along the $t=0$ section the upper part of the Penrose diagram of Fig.~\ref{schwarzschildads} with its Euclidean continuation. The Euclidean path integral determines the state (\ref{hhstate}), which then evolves in Lorentzian time. However, it has been argued \cite{Maldacena:2001kr, Hawking:2005kf} that the Euclidean path integral receives contributions from geometries other than the black hole, and that those may restore unitarity. Said differently, this proposal posits that the state (\ref{hhstate}) neglects contributions of other saddle points and is only part of the full, unitary story.

The general arguments of 
Sec.~\ref{eSbasis}-\ref{eSsup}
indicate that the key to restoring unitarity lies in quantum gravity. Saddle points of the Euclidean path integral pick out semiclassical solutions to the equations of motion, so it is hard to imagine how an extra saddle point may ever repair unitarity. It is not surprising, therefore, that a quantitative examination of CFT correlation functions \cite{Barbon:2003aq, Barbon:2004ce,Kleban:2004rx,Barbon:2005jr} did not find the Poincare recurrences expected in a unitary theory. Furthermore, one expects that a {\it sine qua non} for unitarity is finite $N$, yet the onset of additional saddle points happens via the Hawking-Page transition \cite{Hawking:1982dh}, which persists to infinite $N$. The latter objection was raised and emphasized by the authors of \cite{Iizuka:2008hg}; their model is reviewed in Sec.~\ref{ipmodel} above.

\subsection{The black hole final state}
\label{bhfinalstate}

It has been suggested that unitarity in black hole physics may be restored by imposing a boundary condition at the singularity of the black hole \cite{Horowitz:2003he}. Ordinarily, information loss occurs because particles that fall inside a black hole are eaten by the singularity. On the other hand, outgoing Hawking radiation is accompanied by an infalling counterpart, as can be seen from ordinary energy conservation and from the original derivation \cite{Hawking:1974rv}. One expects that the states of outgoing and infalling Hawking radiation are maximally entangled. The proposal of \cite{Horowitz:2003he} is to impose an appropriate boundary condition at the singularity, which will effectively entangle infalling matter and infalling Hawking radiation. The chain of two entanglements, one between infalling matter and infalling Hawking radiation, and one between infalling and outgoing Hawking radiation, would effectively ensure that no information is lost.

A natural objection to this scenario is that it seems to burden the physics with a teleological ingredient.  It has also been pointed out \cite{Gottesman:2003up} that the feasibility of the scenario depends sensitively on the interactions between infalling matter and Hawking radiation inside the black hole. From the viewpoint of the effective field theory describing the black hole interior away from the singularity, the black hole final state proposal is fine-tuned.

\subsection{Remnants}
\label{secremnants}

We have not yet considered the logical possibility that the process of black hole evaporation in fact emits no information to the environment, but instead leaves behind a remnant that stores the information in its internal state. In addition to its arguable lack of parsimony, this idea suffers from an infinite production problem \cite{Giddings:1994qt}: since the remnant must be able to encode the pedigree of every black hole, its density of states must skyrocket around the Planck scale, which leads to infinite pair production. Evading the problem usually runs into problems with energy conservation, locality, crossing symmetry, or leads to fine-tuned interactions between remnants and ordinary fields.

An interesting realization of the idea of remnants involves a third quantized theory of baby universe remnants \cite{Polchinski:1994zs}. In another scenario, black hole singularities are conjectured to connect two distinct semiclassical spacetimes. It has been argued \cite{Smolin:1994vb} that this conjecture unifies a resolution of the black hole information paradox with a retrodiction of the parameters of the standard models of particle physics and cosmology. This happens via a selection principle loosely analogous to anthropic reasoning, which favors models that maximize the production of black holes. For a recent account of baby universe remnants, see \cite{Hossenfelder:2009xq} and references therein.

\section{Open problems and outlook}
\label{further}

This paper has reviewed many approaches to information recovery from black holes, especially in the context of the AdS/CFT correspondence where the problem can be posed precisely and sometimes quantitatively.  We have emphasized that the key to both information loss and recovery lies in phenomena that correct the observables at $O(e^{-S}) = O(e^{-A/4 G_N \hbar})$.  Such non-perturbatively tiny corrections to observables are inevitable in any theory that microscopically realizes the enormous statistical degeneracy associated to a black hole.

A central question that we have not discussed is how and why the entropy of black holes gets realized as a geometric quantity in spacetime, i.e., the horizon area.  At present there is no answer to this question or even a particularly good approach to answering it, at least within string theory.\footnote{In loop quantum gravity, the horizon is defined in such a way that its area is inevitably proportional to an entropy.}   Put otherwise, if a system acquires a sufficiently large statistical degeneracy, why does it also develop a distinguished co-dimension one surface in spacetime which reflects the logarithm of the degeneracy?      Mathur has suggested that this phenomenon occurs because the underlying bound states that give rise to black hole degeneracy necessarily acquire a large transverse size \cite{Mathur:1997wb}.  Some of the evidence for this was discussed in Sec.~3.2.3 and the references cited there, but even if this is the case we do not have a quantitative argument explaining why the dynamics of string theory produces a horizon proportional to entropy.   

Another important question is whether the horizon is ``really there''. As emphasized in this review, the causal disconnection of an ``interior'' by a horizon is likely to be an artifact of coarse-graining over quantum gravitational details of the spacetime.  Pushing this sort of idea to its logical limits  suggests the possibility, raised by Verlinde,  that gravity might really be an entropic force \cite{Verlinde:2010hp}.   If this point of view can be quantitatively realized, one would presumably find that semiclassical probes are absorbed into black holes and are unable to get out for phase space reasons -- i.e. there are many more configurations with the probe absorbed than released.  While there does not appear to be any room in this picture for the destruction of fine-grained information (and plenty of room for the destruction of coarse-grained information), it would be interesting to better understand how our conventional picture of Hawking radiation, the black hole interior, asymptotic observers, and infalling observers could fit within an entropic gravity scenario.  Is there, for example, a precise notion of black hole complementarity here?

The approaches to information loss discussed in this review largely address the point of view of the asymptotic observer and the recovery of information at infinity.    But what about the perspective of the infalling observer?  In the semiclassical picture such an observer effectively propagates in a well-defined ``interior geometry'' before encountering a singularity.  Does this picture remain valid in a fully quantum theory that resolves information loss, and if so, how?   A recent approach to this question involved the holographic description of D-branes falling through the horizon of an AdS black hole \cite{Horowitz:2009wm}.  The authors found evidence that the fully quantum spacetime should not simply be viewed  as a small smoothing out of the semiclassical geometry near the singularity.  Rather, they suggested that D-brane probes penetrating the horizon effectively enter a non-geometric phase of some kind, so that concepts like a global event horizon that are well defined in the classical limit simply do not have a meaningful quantum analog.  From this perspective, the event horizon and the causal disconnection of an ``interior'' region appear as artifacts of the classical limit (Fig.~4 of \cite{Horowitz:2009wm}) very much as advocated in this review.  It would be very useful to study such scenarios more carefully.

Understanding the black hole interior holographically  would certainly shed light on information loss and recovery in these geometries.  The effort would also elucidate other basic puzzles in gravity.  One of these is whether time can be emergent.    Recall that the radial direction of classical black holes becomes timelike behind the horizon and this is why observers penetrating the horizon are necessarily drawn into the the singularity.  A holographic understanding of the black hole interior must somehow describe this interior time as an emergent phenomenon.   Via the AdS/CFT correspondence we have many precise examples of emergent {\it space}, and there have been occasional suggestions that the time in cosmological settings might emerge from the collective dynamics of a Euclidean theory \cite{Strominger:2001pn,Balasubramanian:2001nb,Balasubramanian:2006sg}.  However, the latter proposals are on much less solid footing than the standard AdS/CFT correspondence.   Related to this is the issue of resolving the spacelike (localized in time) singularities of classical gravity.  Such singularities appear inside Schwarzschild black holes and at the Big Bang.   String theory has provided beautiful resolutions to many kinds of singularities in gravity, but all of these have been timelike (localized in space) or null.  There may be some connection between the challenge of resolving spacelike singularities, the question of emergent time, and the conundrum of information loss in black holes.

The approaches to the information problem reviewed in this paper seek to reconcile unitarity on a fundamental level with information loss on the level of an effective description. This implies that a sufficiently powerful observer can recover information fallen into a black hole. However, there is a trade-off between how fast and how easy this recovery process is: as an example, the conclusion that black holes are information mirrors \cite{Hayden:2007cs} (see Sec.~\ref{secmirror}) relies on the assumption that they eject information in a maximally homogenized form. It is possible that the two-step process of (1) waiting for information to come out of a black hole and (2) decoding the information from a scrambled signal is subject to a fundamental limitation, with a barrier to step~(2) coming from complexity theory. Indeed, it has been suggested \cite{Aaronson:2005qu} that complexity theory may contain lessons about fundamental laws of nature.\footnote{Ref.~\cite{Abrams:1998vz} is an example of an interesting interplay between complexity theory and quantum mechanics.} Black holes are perhaps the most likely arena for realizing this possibility. 

\section*{Acknowledgements}
We thank our collaborators for extensive discussions of the ideas presented in this review.  BC is supported by the Natural Sciences and Engineering Research Council of Canada.  VB is supported by DOE grant DE-FG02-95ER20893.

\end{document}